\documentclass[fleqn,usenatbib]{mnras}

\usepackage{newtxtext,newtxmath}
\usepackage[version=4]{mhchem}

\usepackage[T1]{fontenc}

\DeclareRobustCommand{\VAN}[3]{#2}
\let\VANthebibliography\thebibliography
\def\thebibliography{\DeclareRobustCommand{\VAN}[3]{##3}\VANthebibliography}

\usepackage{graphicx}	
\usepackage{amsmath}	

\title[Formaldehyde and methanol in cold cores]{Correlation between formaldehyde and methanol in prestellar cores}

\author[Anna Punanova et al.]{
A. F. Punanova,$^{1}$\thanks{E-mail: punanova@chalmers.se} 
K. Borshcheva,$^{2,3}$ 
G. S. Fedoseev,$^{4,2,5}$ 
P. Caselli,$^{6}$ 
D. S. Wiebe,$^{3}$ 
A. I. Vasyunin$^{2}$ 
\\
$^{1}$Onsala Space Observatory, Observatoriev\"agen 90, R\aa\"o, 43992 Onsala, Sweden\\
$^{2}$Ural Federal University, 620002, 19 Mira street, Yekaterinburg, Russia\\
$^{3}$Institute of Astronomy of the Russian Academy of Sciences, 119017, 48 Pyatnitskaya street, Moscow, Russia\\
$^{4}$Xinjiang Astronomical Observatory, Chinese Academy of Sciences, Urumqi 830011, China \\
$^{5}$Xinjiang Key Laboratory of Radio Astrophysics, Urumqi 830011, China \\
$^{6}$Max-Planck-Institut f\"ur extraterrestrische Physik, Giessenbachstrasse 1, 85748 Garching, Germany
}

\date{Accepted 30 January 2025. Received 27 January 2025; in original form 3 December 2024}

\pubyear{\the\year{}}

\begin{document}
\label{firstpage}
\pagerange{\pageref{firstpage}--\pageref{lastpage}}
\maketitle

\begin{abstract} 
Formaldehyde is a key precursor in the formation routes of many complex organic molecules (COMs) in space. It is also an intermediate step in CO hydrogenation sequence that leads to methanol formation on the surface of interstellar grains in cold dense prestellar cores where pristine ices are formed. Various chemical models successfully reproduce the COMs abundances in cold cores, however, they consistently overpredict the abundance of formaldehyde by an order of magnitude. This results in an inverse \ce{H2CO}:\ce{CH3OH} abundance ratios obtained in the astrochemical simulations as compared to the observed values. In this work, we present a homogeneous data set of formaldehyde observational maps obtained towards seven dense cores in the L1495 filament with the IRAM 30~m telescope. Resolving the spatial distribution of the molecules is essential to test the chemical models. We carefully estimate the formaldehyde column densities and abundances to put reliable observational constraints on the chemical models of cold cores. Through numerous tests, we aim to constrain the updated chemical model MONACO to better align with the observed formaldehyde abundance and its ratio to methanol. In particular, we elaborate on the branching ratio of the \ce{CH3} + O reaction at low temperatures. The revised MONACO model reproduces abundances of both methanol and formaldehyde within an order of magnitude. However the model tends to overproduce formaldehyde and underpredict methanol. Consequently, the model systematically overestimates the \ce{H2CO}:\ce{CH3OH} ratio, although it remains within an order of magnitude of the values derived from observations.
\end{abstract}




\section{Introduction} \label{sec:intro}

Formaldehyde (\ce{H2CO}) is widely observed in star-forming regions and is detected in interstellar ices \citep{Boogert2008,Yang2022}. Along with its direct chemical derivatives, HCO, \ce{CH3O} and \ce{CH2OH}, it plays a central role in the cold formation routes of various O-bearing complex organics molecules (COMs), including aldehydes and polyalcohols \citep[see, e.g.,][]{Chuang2016,Fedoseev2017}. COMs are now detected towards many cold pre-stellar cores \citep[e.g. L1544, B5, cores in L1495, L1521E;][]{Vastel2014,Jimenez-Serra2016,Taquet2017,Scibelli2020,Scibelli2021}, which are the initial stage of star formation and a convenient laboratory to explore the primordial formation of COMs. In such environment ($T\simeq10$~K, $n>10^4$~cm$^{-3}$), formaldehyde formation is directly connected to formation of methanol, which is often referred to as the simplest COM and used as the abundance reference for COMs in observational and theoretical studies \citep[e.g.,][]{Jimenez-Serra2016,Vasyunin2017,Scibelli2021}. 

While models successfully reproduce the observed abundances of CO, methanol and COMs, they overpredict the gas-phase abundance of formaldehyde, including its abundance with respect to that of methanol. The models predict higher gas-phase abundance of formaldehyde compared to methanol, towards the cold cores \citep[e.g.,][]{Vasyunin2017,Sipila2020,Chen2022,Garrod2022,Potapov2024}. This contradicts the observational results that show an opposite trend \citep[e.g.,][]{Guzman2013,Cuadrado2017,Chacon-Tanarro2019a,Kirsanova2021,Mercimek2022,Freeman2023}. To the best of our knowledge only two chemical models \citep{Garrod2007,Garrod2022} showed the prevalence of methanol over formaldehyde at temperatures (8--12~K) and the chemical ages ($\sim$500--900~kyr) typical of cold dense cores, however with no attempts to reproduce observed CO depletion or molecular distribution profiles across the cores.

In prestellar cores, methanol is supposed to be formed on the surface of dust icy mantles via subsequent hydrogenation of CO \citep[e.g.,][]{Hiraoka1994,Watanabe2002,Fuchs_ea09}. Formaldehyde is an intermediate product of this reaction sequence (``ladder''). At each step of the ``ladder'', reaction products may desorb to the gas phase and become available for sub-millimetre observations. Also each of the species produced in this hydrogenation ``ladder'' ($ \rm CO \leftrightarrow HCO \leftrightarrow H_2CO\leftrightarrow CH_2OH / CH_3O \leftrightarrow CH_3OH$) can participate in the H-atom abstraction reactions, resulting in reformation of the initial reactant \citep[see, e.g.,][]{Hidaka2009,Minissale2016,Chuang2018}. The prevalence of methanol may indicate that the balance at the last step is shifted in favour of methanol formation or its desorption probability. The \ce{H2CO}:\ce{CH3OH} ratios can provide valuable hints for the overall balance in the CO hydrogenation sequence or the reactive desorption probabilities at its steps. Recent laboratory studies demonstrated that at the last step of this reaction ``ladder'', CH$_3$O reacts with H$_2$ or H$_2$CO instead of atomic H, to form CH$_3$OH \citep{Santos2022}. This may also affect the ``equilibrium'' between the \ce{CH3OH} formation and destruction paths, as well as \ce{CH3OH} reactive desorption probability. In addition to the surface formation route, formaldehyde can also form in the gas phase via the reaction O + CH$_3$ $\rightarrow$ H$_2$CO + H \citep[e.g.,][]{Baulch2005,Hack2005,Xu2015}. 

In this work, we put under the test formation and destruction routes for formaldehyde employed in the majority of chemical models of cold cores, as well as the mechanisms of adsorption (freeze-out) and desorption. We focus on the missing steps, which absence  may lead to the overproduction of formaldehyde. We benchmark our findings with a large and homogeneous observational set of formaldehyde (presented in this work) and methanol lines \citep[presented in][]{Punanova2022} towards seven dense cores in the L1495 filament in Taurus, a nearby \citep[130--135~pc distant;][]{Schlafly2014,Roccatagliata2020} site of low-mass clustered star formation. We present the observational maps of formaldehyde column density and relative abundance with respect to \ce{H2} (hereinafter -- abundance). We discuss the correlation between methanol and formaldehyde abundance in cold cores, the problems with the modelling of the formaldehyde chemistry and possible reasons for the observed low abundance of formaldehyde in star-forming regions. 

This paper is structured in the following way. In Sect.~\ref{sec:obs} we describe the observations and data reduction. In Sect.~\ref{sec:results} we briefly describe the results from our previous works that we use, present our new observational results, and the results of our astrochemical simulations. In Sect.~\ref{sec:discussion} we discuss the implications of our findings, and in Sect.~\ref{sec:conclusions} we give our conclusions.

\section{Observations and Data Reduction} \label{sec:obs}

\begin{table}
\caption{The observed cores. The numbers are given according to \citet{Hacar2013}, H2013, for cores 1, 6, 7, 10, 11, 16, and \citet{Seo2015}, S2015, for core 35. The given coordinates are the central positions of the maps. The only protostellar core is indicated with an asterisk (*). The region names are given according to \citet{Barnard1927}.}\label{Tab:sources}
\centering
\begin{tabular}{lcccr}
\hline
\multicolumn{2}{c}{Core} & $\alpha_{J2000}$ & $\delta_{J2000}$ & Region \\
H2013 & S2015 & ($^h$ $^m$ $^s$) & ($^{\circ}$ $^{\prime}$ $^{\prime\prime}$) & \\
\hline
1 & 12, 13, 14 & 04:17:42.347 & 28:07:30.88 & B10 \\
6 & 8, 9 & 04:18:06.379 & 28:05:34.87 & B10 \\
7 & 20 & 04:18:11.343 & 27:35:33.07 & B211 \\
10 & 22 & 04:19:36.768 & 27:15:32.00 & B213 \\
11* & 23* & 04:19:42.154 & 27:13:31.03 & B213 \\
16 & 33 & 04:21:20.595 & 27:00:13.63 & B213 \\
-- & 35 & 04:24:20.600 & 26:36:02.00 & B216 \\
\hline
\end{tabular}
\end{table}

\begin{table*}
\caption{The observed formaldehyde lines.}\label{Tab:lines}      
\centering                          
\begin{tabular}{lccccccccr}       
\hline                
Transition & Frequency & $E_{\rm up}/k$ & $A$ & $C$ & $F_{\rm eff}$ & $B_{\rm eff}$ & $\Delta \varv_{\rm res}$ & rms in $T_{\rm mb}$ & $T_{\rm sys}$\\  
 & (GHz) & (K) & ($10^{-5}$~s$^{-1}$) & (cm$^{3}$~s$^{-1}$) & & & (km~s$^{-1}$) & (K) & (K) \\   
\hline                        
ortho-(2$_{1,2}$--$1_{1,1}$) & 140.839502 & 21.92 & 5.3040 & 7.8$\times10^{-11}$ & 0.93 & 0.74 & 0.11 & 0.08--0.17 & 100--160 \\
 para-(2$_{0,2}$--$1_{0,1}$) & 145.602949 & 10.48 & 7.8130 & 4.7$\times10^{-11}$ & 0.93 & 0.73 & 0.10 & 0.08--0.13 & 100--160 \\
ortho-(2$_{1,1}$--$1_{1,0}$) & 150.498334 & 22.62 & 6.4720 & 6.2$\times10^{-11}$ & 0.93 & 0.72 & 0.10 & 0.08--0.25 & 100--160 \\
\hline                                   
\end{tabular}\\
\begin{flushleft}
{\bf Notes.}
The frequencies and energies for the \ce{H2CO} lines are taken from \citet{Bocquet1996}, accessed through the JPL database \citep{Pickett1998}. The Einstein coefficients, $A$, and the collisional coefficients, $C$, are taken from the LAMDA database \citep{Schoeier2005}. The collisional coefficients in the LAMDA database were calculated based on \citet{Wisenfeld2013}. 
\end{flushleft}
\end{table*}

We present here the maps of three formaldehyde lines (ortho-\ce{H2CO} at 140.8 and 150.5~GHz and para-\ce{H2CO} at 145.6~GHz) towards seven dense cores of the L1495 filamentary structure (see Table~\ref{Tab:sources}). We mapped the formaldehyde line at 145.6~GHz with the IRAM 30~m telescope (IRAM projects 013-18, 125-18, and 031-19) simultaneously with four methanol lines at 96.7~GHz and 145.1~GHz \citep[the methanol observations are presented in][]{Punanova2022}. The observations were done on 17--23 October 2018, 27--29 March and 16 September 2019 under fair weather conditions, pwv=1--10~mm. The on-the-fly formaldehyde maps were obtained with the EMIR 150 (2~mm band) heterodyne receiver\footnote{\url{http://www.iram.es/IRAMES/mainWiki/EmirforAstronomers}} in position switching mode, using the FTS~50 backend. The spectral resolution of the FTS~50 data was 50~kHz, the corresponding velocity resolution was 0.10~km~s$^{-1}$. The other two formaldehyde lines at 140.8~GHz and 150.5~GHz were taken from the FTS data of the previous IRAM projects \citep[032-14, 156-14, described in][]{Punanova2018b}. There are no 150.5~GHz observations for one of the sources, core~35, since it was not included in the previous projects. The exact line frequencies, beam efficiencies, beam sizes, spectral resolutions and sensitivities are given in Table~\ref{Tab:lines}. In all observing runs, sky calibrations were obtained every 10--15 minutes. Reference positions were chosen individually for each core to make sure that the positions were free of any methanol and formaldehyde emission. Pointing was checked by observing QSO~B0316+413, QSO~B0439+360, QSO~B0605-085, Uranus, Mars, or Venus every two hours and focus was checked by observing QSO~B0439+360, Uranus, Mars, or Venus every six hours.   

All methanol and formaldehyde maps were convolved to the 26.8$^{\prime\prime}$ beam and 8.7$^{\prime\prime}$ pixel size, that fulfils the Nyquist sampling criterion, to make the maps directly comparable. The convolved spectral cubes were created with the native IRAM software GILDAS/CLASS\footnote{Continuum and Line Analysis Single-Dish Software \url{http://www.iram.fr/IRAMFR/GILDAS}.}. The spectral line analysis was performed with the Pyspeckit module of Python \citep{Ginsburg2011}.

In this work, we use the molecular hydrogen column density $N_{\rm tot}$(H$_2$), dust temperature $T_{\rm dust}$ and visual extinction $A_V$ measured via dust continuum emission observations done with Herschel/SPIRE\footnote{Herschel is an ESA space observatory with science instruments provided by European-led Principial Investigator consortia and with important participation of NASA.} \citep{Palmeirim2013}, as well as the column density of CO and CO depletion factor $f_d$, calculated and presented in \citet{Punanova2022}, based on the data, presented in \citet{Tafalla2015}. 

\section{Results} \label{sec:results}

\subsection{Formaldehyde distribution}

\begin{figure*}
    \centering
    \includegraphics[height=8.5cm,keepaspectratio]{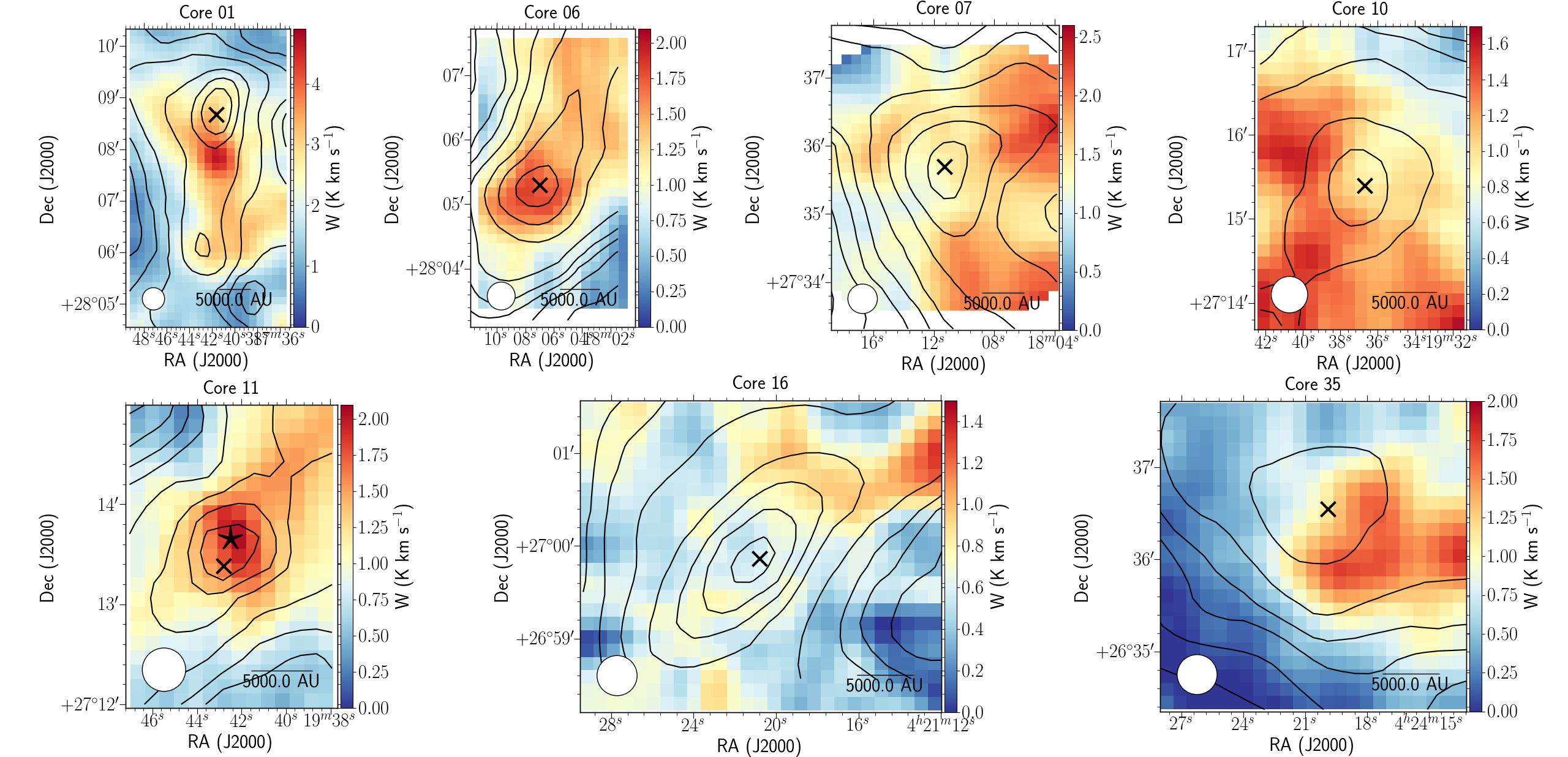}
    \caption{Integrated intensity of all observed formaldehyde lines toward the observed cores (colour scale) and visual extinction (black contours at $A_V$~= 3, 4, 5, 6, 8, 12, 16, 20, and 24~mag). The top $A_V$ contours are at $A_V$~= 24~mag for core 16 and at $A_V$~= 20~mag for the other cores. The black star shows the position of Class~0 protostar IRAS 04166+2706 \citep{Santiago-Garcia2009}; crosses show the Herschel/SPIRE dust emission peaks. The white circle at the bottom left of each map shows the 26$^{\prime\prime}$ beam of the maps.}
    \label{fig:form-int-int-maps}
\end{figure*}

Figure~\ref{fig:form-int-int-maps} shows the sum integrated intensities of the three observed formaldehyde lines and contours of visual extinction. Cores 6 and 11 show the highest formaldehyde intensity towards the dust peaks; core~1 also shows the formaldehyde intensity peak towards its centre, but offset from the dust peak; core 35 shows the formaldehyde emission peak as a quite vast patchy spot also away from the dust peak. While cores 1, 6, 11, and 35 show concentrated formaldehyde emission, cores 7, 10, and 16 show no distinct pattern in their formaldehyde emission except for the fact that it is depleted towards the dust peaks. In cold cores, formaldehyde is formed both in the gas phase and on the surface and in the bulk of dust icy mantles. In the latter case, it can further desorb to the gas phase. Unlike methanol, it should not necessarily be abundant towards the zone of heavy freeze-out \citep[e.g.,][]{Vasyunin2017,Punanova2022} due to the gas phase formation mechanism. 


\begin{figure*}
    \centering
    \includegraphics[height=8.5cm,keepaspectratio]{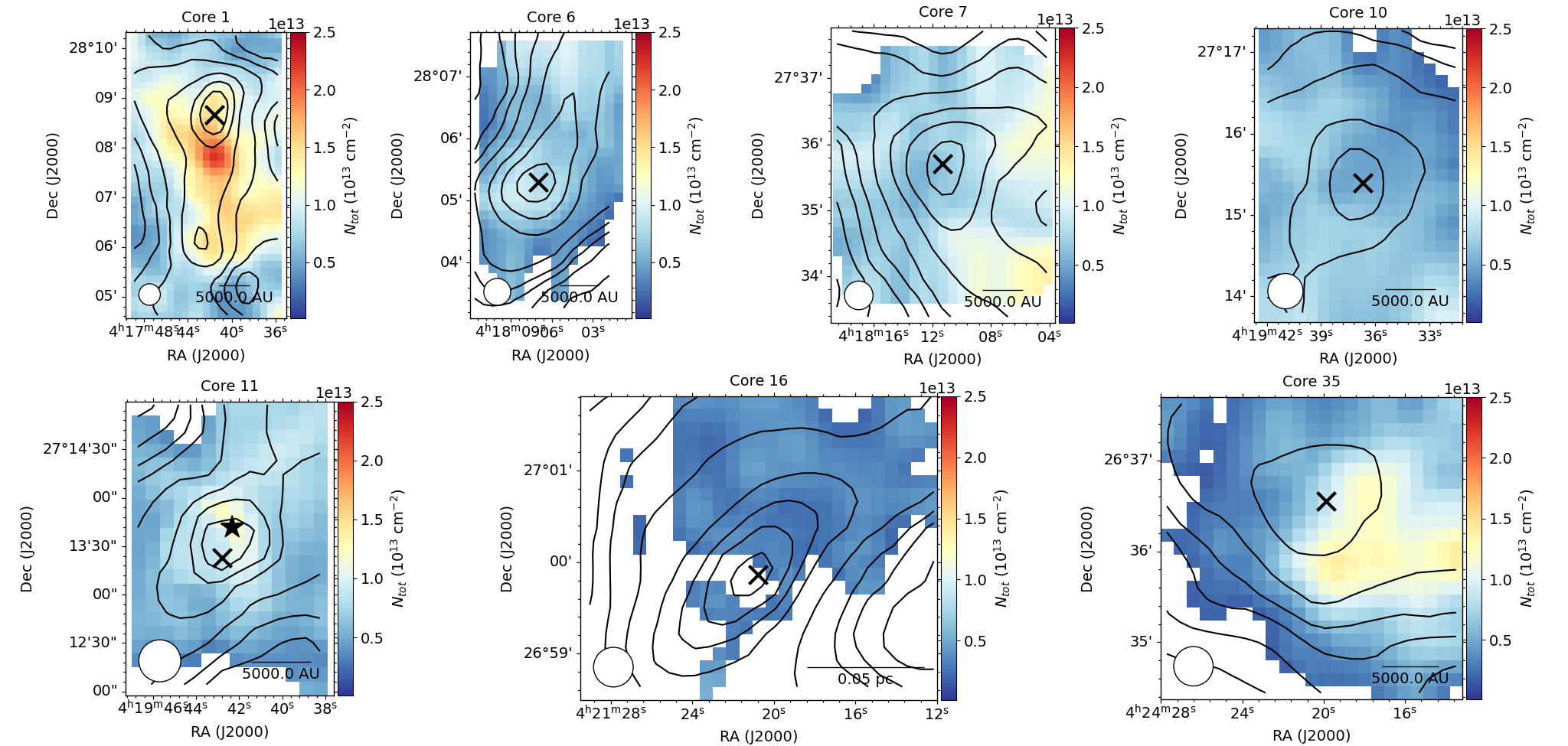}
    \caption{Column density of ortho-\ce{H2CO} measured via the 140~GHz line toward the observed cores (colour scale) and visual extinction (black contours). The contours and other symbols are drawn like in Fig.~\ref{fig:form-int-int-maps}. The same colour scale is used in all the panels to emphasize the differences between the cores.}
    \label{fig:form-140-Ntot-maps}
\end{figure*}

\begin{figure*}
    \centering
    \includegraphics[height=8.5cm,keepaspectratio]{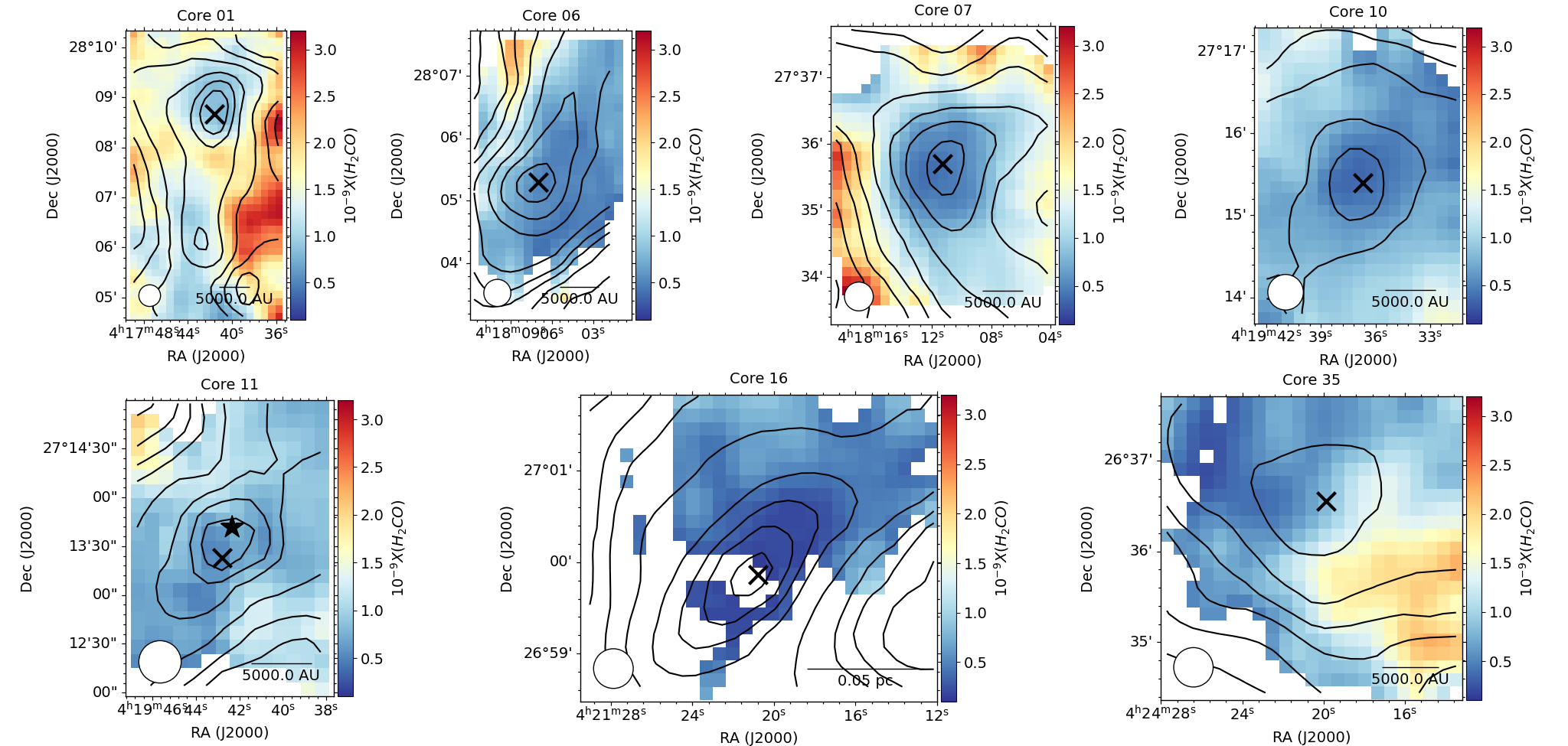}
    \caption{The total abundance of \ce{H2CO} measured via the 140~GHz line with an assumption of o/p=2.0 toward the observed cores (colour scale) and visual extinction (black contours). The contours and other symbols are drawn like in Fig.~\ref{fig:form-int-int-maps}. The same colour scale is used in all the panels to emphasize the differences between the cores.}
    \label{fig:x-form-maps}
\end{figure*}

\subsection{Formaldehyde column density and abundance}

In our observational set, we have two ortho-H$_2$CO lines with close energies of the upper level and one para-H$_2$CO line. Even assuming the o/p ratio =~1 we could not build reliable rotational diagrams. Thus we consider the ortho- and para- variants of the molecule as different species. We estimate the formaldehyde column densities with two different methods: with the assumption of optically thin lines consistent with LTE and also make the column density profiles across our cores via non-LTE modelling with RADEX \citep{vanderTak2007} applying the physical structure designed for our chemical model (see Sect.~\ref{sec:phys_mod} for details) and assuming o/p ratio of \ce{H2} of 10$^{-4}$ \citep[e.g.,][]{Sipila2013,Furuya2015}. 

The simple assumption of LTE, optically thin lines, and excitation temperature of 10~K yields very low column densities of para-H$_2$CO, 2--15 times smaller than that of ortho-H$_2$CO while the o/p ratio of \ce{H2CO} can take values up to 3:1 \citep[e.g.,][]{Kahane1984}. RADEX modelling gives o/p=0.9--2.8 of \ce{H2CO} towards the dust peaks, with an average value of 2.0. The formaldehyde formation temperature of 10~K corresponds to the 1:1 o/p ratio \citep{Kahane1984}. However, the observations of cold clouds give o/p=1--2 \citep[e.g.,][]{Kahane1984,Mangum1993}. Besides that, \citet{Yocum2023} could not reproduce the 1:1 o/p ratio of gas phase \ce{H2CO} with the ice formation temperature of 10~K in their laboratory work. They obtained o/p=3:1 regardless of the ice temperature of 10--40 K and suggested that lower o/p ratio indicated the gas-phase \ce{H2CO} formation. The column density of ortho-H$_2$CO estimated with RADEX is close to that measured under the assumption of LTE and optically thin lines. It is about a factor of $\sim$2 lower towards the dust peak (see Table~\ref{tab:col_den}) and higher towards the core edges. Since the observed cores are not spherical and the number of the formaldehyde lines and their close energies do not allow us to estimate simultaneously the gas density and the column density via RADEX across all cores, we use the LTE column densities measured through the 140~GHz ortho-\ce{H2CO} line (shown in Fig.~\ref{fig:form-140-Ntot-maps}) and use them further in the analysis, adopting the average o/p ratio =~2.0 to calculate the total \ce{H2CO} abundance. The ambiguity of the observed o/p ratio of \ce{H2CO} introduces a factor of 2 uncertainty in our $N_{\rm tot}$(\ce{H2CO}). This would still result in a low abundance of formaldehyde with respect to methanol.

\begin{table*}
    \caption{Formaldehyde column densities towards the dust peaks estimated via non-LTE RADEX and under assumption of LTE and optically thin lines. The collisional coefficients between o-/p-\ce{H2CO} and o-/p-\ce{H2} for RADEX are taken from \citet{Wisenfeld2013} through LAMDA database \citep{Schoeier2005}.}
    \label{tab:col_den}
    \centering
    \begin{tabular}{lccc|ccc|ccc|cc}
    \hline
     & & & & \multicolumn{3}{c}{RADEX o-\ce{H2CO}} & \multicolumn{3}{c}{RADEX p-\ce{H2CO}} & LTE o-\ce{H2CO} & LTE p-\ce{H2CO} \\
    Core & $n$(o-\ce{H2}) & $n$(p-\ce{H2}) & $T_k$ & $T_{\rm mb}$ & $\varv$ & $N_{\rm tot}$ & $T_{\rm mb}$ & $\varv$  & $N_{\rm tot}$ & $N_{\rm tot}$ & $N_{\rm tot}$ \\
     & (cm$^{-3}$) & (10$^5$ cm$^{-3}$) & (K) & (K) & (km s$^{-1}$) & (10$^{12}$ cm$^{-2}$) & (K) & (km s$^{-1}$) & (10$^{12}$ cm$^{-2}$) & (10$^{12}$ cm$^{-2}$) & (10$^{12}$ cm$^{-2}$) \\
     \hline
    1  & 12.4 & 1.24 & 10.27 & 1.509 & 0.648 & 10.08 & 1.124 & 0.652 & 4.63 & 15.15 & 2.67 \\
    6  & 12.1 & 1.21 & 10.19 & 1.571 & 0.318 & 5.39 & 0.849 & 0.556 &  2.86 & 8.38 & 1.44 \\
    7  & 7.4  & 0.74 & 10.37 & 0.550 & 0.787 & 4.94 & 0.728 & 0.520 &  3.28 & 5.89 & 1.20 \\
    10 & 7.8  & 0.78 & 10.71 & 0.724 & 0.447 & 3.61 & 0.500 & 0.377 &  1.40 & 5.79 & 0.77 \\
    11 & 24.3 & 2.43 & 11.42 & 0.954 & 0.641 & 3.55 & 0.730 & 0.523 &  1.27 & 8.65 & 1.04 \\
    16 & 12.3 & 1.23 &  9.76 & 0.382 & 0.414 & 1.31 & 0.489 & 0.490 &  1.39 & 2.46 & 0.84 \\
    35 & 5.4  & 0.54 & 10.76 & 0.950 & 0.800 & 12.45 & 0.802 & 0.636 & 5.79 & 9.67 & 1.44 \\

    \hline     
    \end{tabular}
\end{table*}

The column densities of ortho-\ce{H2CO} vary in the range 0.1--2.5$\times$10$^{13}$~cm$^{-2}$, with higher column density observed in the proximity of the dust peaks in cores 1, 6, 11 and towards the shells in cores 7, 10, 16, and 35 (see Fig.~\ref{fig:form-140-Ntot-maps}). This distribution is similar to that of methanol with the exception of core 1, where methanol showed a distinct ring-like structure around the core with a bright intensity blob on a side \cite[for the methanol maps, see Fig.~17 of][]{Punanova2022}.

To obtain the total formaldehyde abundance, we divided our ortho-H$_2$CO column densities by $N_{\rm tot}$(H$_2$) from \citet{Punanova2022} and apply a factor to account for the o/p =~2.0. The abundance varies in the range 0.1--3.2$\times10^{-9}$ with a median value of 1.1$\times10^{-9}$. The maps of total formaldehyde abundance (see Fig.~\ref{fig:x-form-maps}) show the highest abundance towards the core shells and prominent depletion towards the dust peaks, just like methanol abundance \citep[Fig.~6 of][]{Punanova2022}. Among the cores, the highest formaldehyde abundance is observed towards core 1 (also with the highest median abundance of 1.5$\times10^{-9}$), where the highest methanol abundance is also observed. However, the \ce{H2CO} abundance peak does not match with the \ce{CH3OH} abundance peak although is situated next to it at an almost the same $A_V$ level. Interestingly, there is a secondary \ce{H2CO} abundance peak almost towards the dust peak of core 1 (where the $N$(\ce{H2CO}) peak is located). There is no methanol enhancement towards that spot. The similar pattern -- \ce{H2CO} abundance peak next to the \ce{CH3OH} peak at the same $A_V$ level and the secondary \ce{H2CO} peak close to the dust peak where \ce{CH3OH} is depleted -- was observed towards the prototypical prestellar core L1544 \citep{Chacon-Tanarro2019a}. This might indicate active formaldehyde formation in the gas phase or a locally enhanced desorbtion.  

\subsection{Chemical modelling}

\subsubsection{Physical profiles}\label{sec:phys_mod}

\begin{figure*}
     \centering
     \includegraphics[height=7.5cm,keepaspectratio]{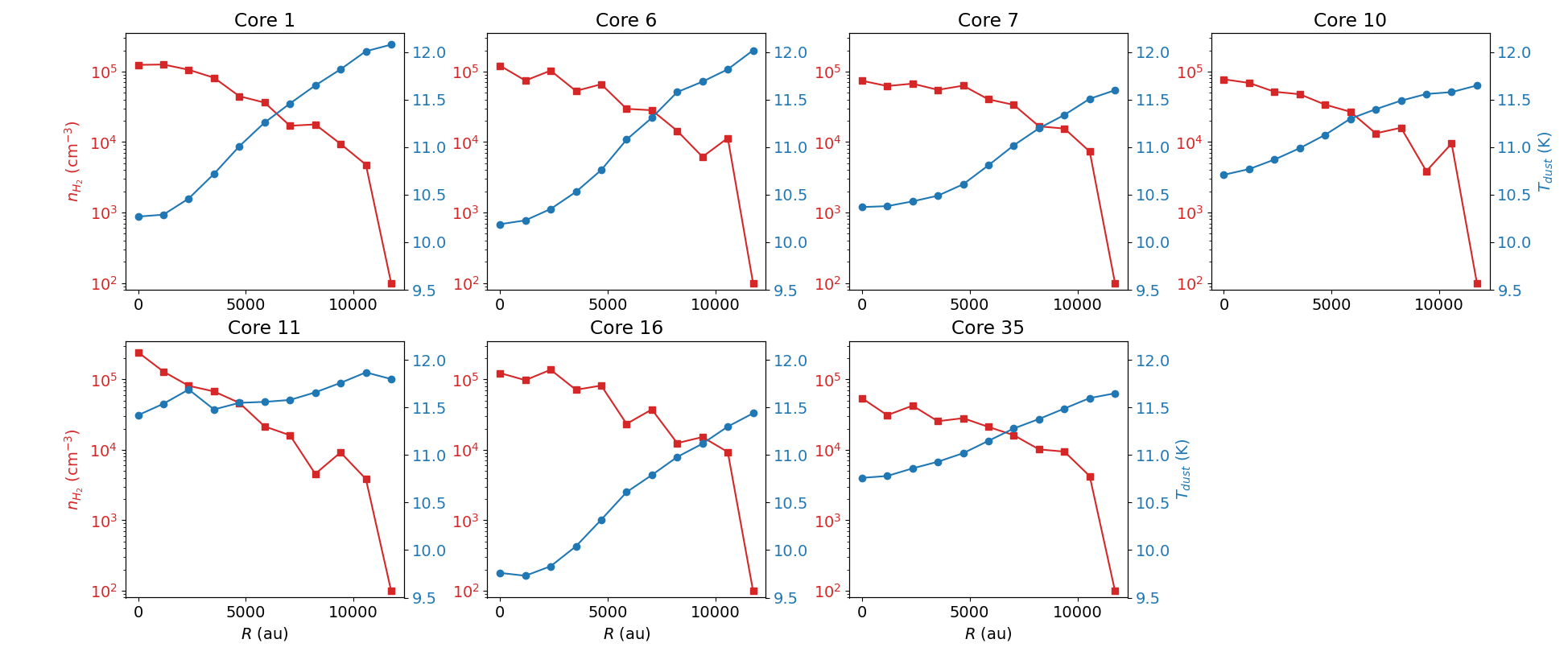}
     \caption{Physical models of the cores: molecular hydrogen number density (red) and dust temperature (blue).}
     \label{fig:phys_profiles}
 \end{figure*}

The physical profiles are built in a similar manner to that of \citet{Punanova2022}. We use the $N$(\ce{H2}) and $T_{\rm dust}$ estimated through the dust continuum emission observations from Herschel/SPIRE (see Sect.~\ref{sec:obs}) and presented in \citet{Punanova2022}. Instead of using polynomial fits to radial distributions of $N$(\ce{H2}) and $T_{\rm dust}$, we use an average over a ring of one pixel width, centred at the dust peak. Ten concentric rings are sufficient to describe the cores, where the outer values of $N$(\ce{H2}) are similar to each other ($\simeq 9 \cdot 10^{21}$~cm$^{-2}$). Then $N$(\ce{H2}) is converted to $n(\ce{H2})$ under an assumption of spherical symmetry of the cores, with the same structure on the line of sight and in the plane of sky, as described in \citet{HasenbergerAlves21} and applied in \citet{Punanova2022}. The gas temperature is considered equal to $T_{\rm dust}$. The profiles are shown in Fig.~\ref{fig:phys_profiles}. This approach compared to the polynomial fit allows to better balance the profiles of the cores with asymmetric structure (where distribution of $N$(\ce{H2}) and $T_{\rm dust}$ is far from circular symmetry). We run the chemical model both with these profiles and with the profiles from \citet{Punanova2022} to be able to compare the new model results to that of \citet{Punanova2022}.

\subsubsection{Model description}

We employ the updated MONACO astrochemical code, which is described in details in Borshcheva et al. (under rev. in ApJ). 
This is a 0D three-phase code based on the rate equations approach. Gas phase, surface ice monolayers and bulk of icy mantles are considered separately, with reactions both on the surface and in the bulk of the ice. Being a 0D code, MONACO produces time-dependent fractional abundances of chemical species independently at every radial point of a physical model of a core. Computed fractional abundances in a set of radial points along the modelled profile are than converted into column densities in order to perform more direct comparison with observations. The key differences of the current code from the MONACO version described in \cite{Vasyunin2017} and utilised in \cite{Punanova2022} are the implementation of non-diffusive processes on dust grains \citep{Jin2020}, which allows adsorbed heavy radicals to react effectively even at the low temperatures typical for prestellar cores, adding H-atom abstraction reactions on dust grains for the CO hydrogenation ``ladder'', correction of H and $\rm H_2$ desorption energy according to the fraction of dust surface covered by $\rm H_2$ \citep{Cuppen2009,GarrodPauly2011} and several new routes to the formation of important ice constituents such as acetaldehyde \citep{Fedoseev2022} and methane \citep{Lamberts2022}.

The chemical evolution of the cores embedded in the L1495 filament is modelled in two steps. The first stage is somewhat different from the one considered in \cite{Vasyunin2017} and \cite{Punanova2022}. As in Borshcheva et al. (under rev. in ApJ), 
the translucent medium employed for the first phase is described with gas density of $10^3$ cm$^{-3}$, visual extinction $A_V = 2.0$~mag and the temperature of gas and dust linearly falling from 15~K to 10~K during the $10^6$~years of evolution. The original chemical composition is the ``low-metal'' abundances from Table~1 in \cite{Wakelam2008}, and hydrogen initially resides in molecular form. The fractional abundances obtained at the final time moment of the first stage are then utilised as the initial chemical composition for each radial point in the second stage where chemistry in the dense cores is simulated. 

Desorption processes incorporated in our modelling are thermal evaporation, cosmic ray-induced desorption \citep{HasegawaHerbst93}, the photodesorption by cosmic-ray-induced UV photons \citep{PrasadTarafdar1983}, reactive (chemical) desorption and photodesorption. The intact photodesorption yield per incident photon equals to $10^{-2}$ for CO \citep{Fayolle2011} and $10^{-5}$ for other species \citep{Cruz-Diaz2016, Bertin2016}; see \cite{Punanova2022} for details. The parametrization of reactive desorption is the same as in Borshcheva et al. (under rev. in ApJ) 
and essentially agrees with the parametrization suggested by \cite{Garrod2007}, which is based on Rice-Ramsperger-Kessel (RRK) theory. The probability of reactive desorption, which is the parameter $a$ in the Expression (2) in \cite{Garrod2007}, is equal to 0.01. The only difference from \cite{Garrod2007} is that the desorption occurs only from the surface fraction not covered by water ice, as discussed in \cite{Vasyunin2017}.

In our current model, the tunnelling through diffusive barriers for light species (H and $\rm H_2$) in the solid phase is switched off, while the tunnelling through reaction activation barriers is switched on. The reaction/diffusion competition for the reactions with a barrier is accounted for \citep{GarrodPauly2011}. The \ce{H2} cosmic-ray ionization rate is $1.3 \times 10^{-17}$~s$^{-1}$. The grain size is $10^{-5}$~cm, the dust-to-gas mass ratio equals 0.01, the dust grain density is 3~g/cm$^3$ and the surface site density equals $1.5 \times 10^{15}$~cm$^{-2}$. The ratio of the diffusion energy barrier of a species to its desorption energy barrier equals 0.5 for atomic species and 0.3 for molecular species. As an example, for H atoms with the desorption energy of 450~K \citep{Minissale2022}, the diffusion barrier is 225~K, which is within the range obtained by the laboratory studies for H diffusion on amorphous solid water \citep{Hama2012} and close to the lowest value of 255~K measured for H diffusion on pure solid CO by \cite{Kimura2018}. The modelled grain mantles predominantly consist of water ice. The ratio of the bulk swapping energy of a species to its diffusion energy in surface monolayers is 1.5 for atomic hydrogen and 2.0 for all other ice constituents. With this ratio, the species heavier than atomic or molecular hydrogen are practically immobile at the low temperatures of prestellar cores, and the diffusive reactions involving only the components heavier than H or $\rm H_2$ have negligible impact on abundances. Four upper ice monolayers are considered as surface \citep{Vasyunin2013}, while the deeper layers of icy mantles constitute bulk ice.

New reactions are added to our chemical network following recent theoretical and experimental studies. As it was mentioned before, hydrogen abstraction reactions are now switched on for every step of the CO hydrogenation sequence \citep{Jin2020}. We also include the experimentally confirmed~\citep[][]{Santos2022} reaction $\rm gCH_3O + gH_2CO \rightarrow gCH_3OH + gHCO$ with a barrier of 2670~K from \cite{Alvarez-Barcia2018}. Prefix ``g'' is for species in the surface layers of icy mantles. When speaking about surface reactions, we imply that we have similar reactions for bulk species in our model. The reaction $\rm gC + gH_2O \rightarrow gH_2CO$, which was shown to produce $\rm H_2CO$ in the solid state \citep{Molpeceres2021}, is present in our network too. For more detailed information on the changes in the chemical network, see Borshcheva et al. (under rev. in ApJ). 

With this model, we investigate the factors which possibly affect the gas-phase $\rm H_2CO$ abundance and the \ce{H2CO}:\ce{CH3OH} ratio. We present the tests that did not result in significant decrease of \ce{H2CO} abundance and \ce{H2CO}:\ce{CH3OH} ratio in the Appendix~\ref{Appendix}. Below, we present the results of the two versions of the model, the default one and the one with an additional channel for the reaction \ce{CH3} + O that is the most effective in $\rm H_2CO$ production.


\subsubsection{Modelling results}
\begin{figure*}
    \centering
    \includegraphics[height=6.0cm,keepaspectratio]{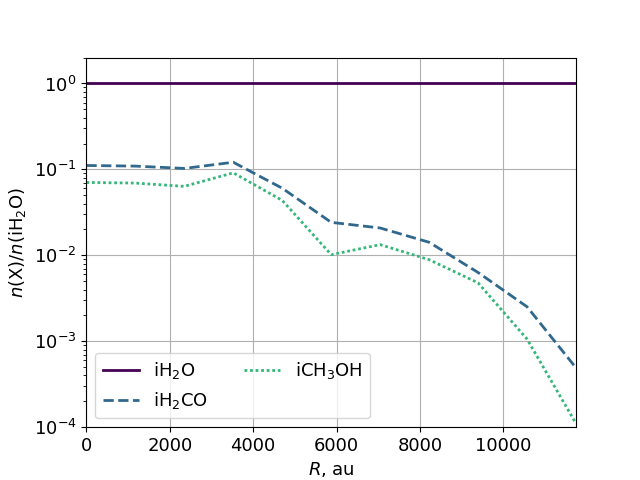}
    \includegraphics[height=6.0cm,keepaspectratio]{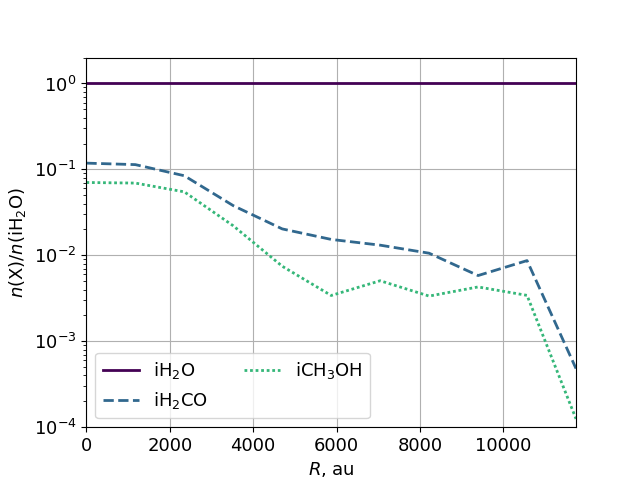}
    \caption{The radial profiles of $\rm H_2CO$ and $\rm CH_3OH$ modelled ice abundances w.r.t. $\rm H_2O$ ice for core 1 (left) and core 10 (right). Prefix ``i'' denotes total amount of ice species, both on surface and in the bulk of ice.}
    \label{fig:H2CO_CH3OH_ice}
\end{figure*}

\begin{figure*}
    \centering
    \includegraphics[height=17.5cm,keepaspectratio]{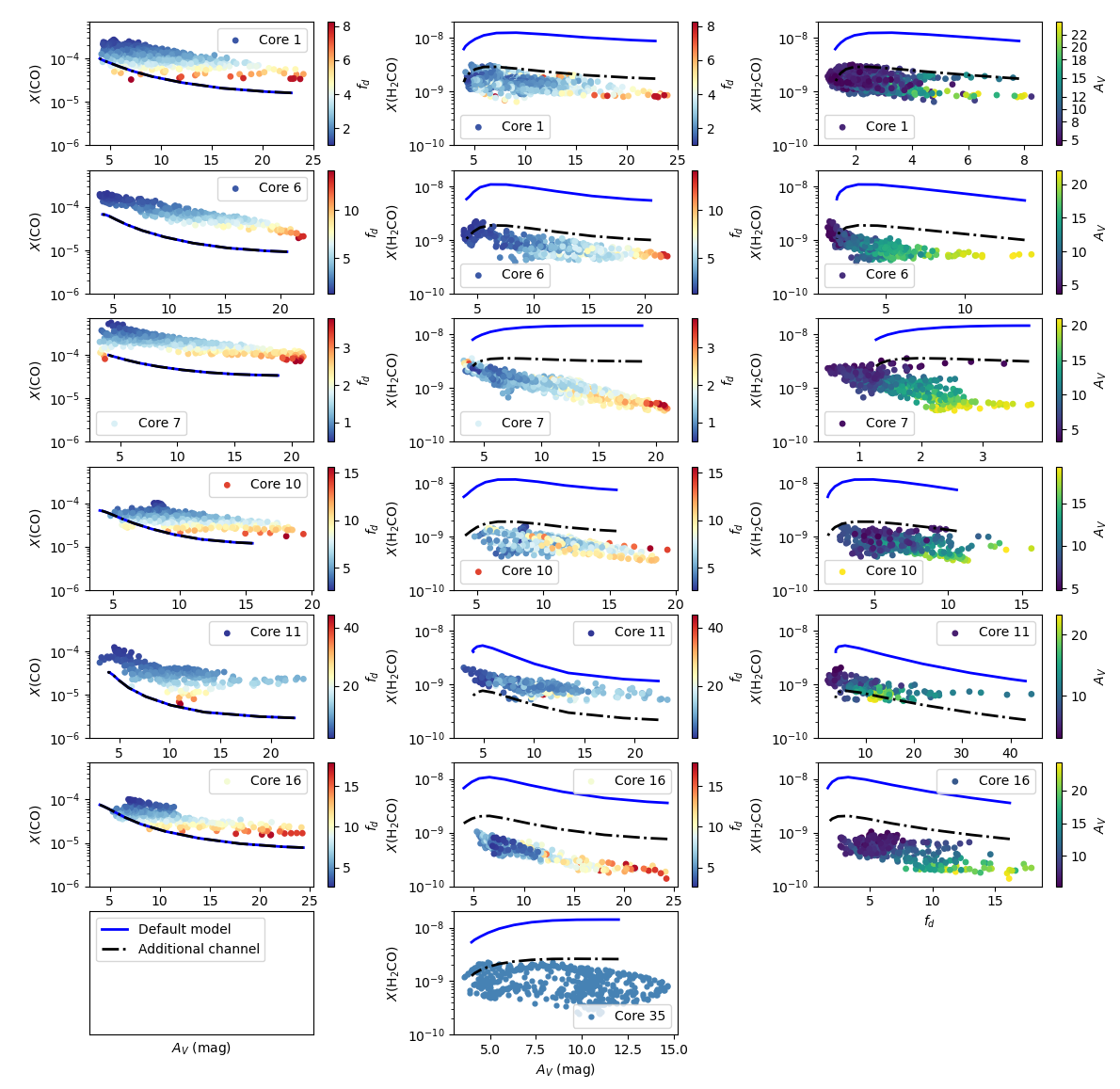}
    \caption{The comparison of the observed (coloured dots) and modelled CO abundance profiles as a function of visual extinction (left); H$_2$CO abundance profiles as a function of visual extinction (middle); and CO depletion factor (right) for all cores. The colour scale represents CO depletion factor $f_d$ (left and middle) and visual extinction $A_V$ (right) in the individual pixels. The data points of core 35 are all shown with the same colour since there are no CO data for this core. The models include the default non-diffusive model (solid blue line) and the non-diffusive model with the additional channel for the reaction \ce{CH3} + O (dashed-dotted black line).
    }
    \label{fig:all-models-and-obs}
\end{figure*}

\begin{table}
    \centering
    \caption{Cores chemical ages corresponding to the CO depletion factor from \citet{Punanova2022}, derived from the chemical model. The age is the same in the default model and in the model with the additional channel for the $\rm CH_3 + O$ reaction. Age 1 shows the age with the physical profiles introduced in this work and Age 2 shows the age with the physical profiles from \citet{Punanova2022}.}\label{tab:fd_age}
    \begin{tabular}{cccc}
    \hline\hline
     Core & $f_d$ & Age 1 & Age 2 \\
         & & (kyr) & (kyr)\\
    \hline
    1  & 8  & 110 & 110 \\
    6  & 14 & 196 & 196 \\
    7  & 4  & 87 & 87 \\
    10 & 12 & 248 & 313 \\
    11 & 44 & 498 & 394 \\
    16 & 17 & 196 & 221 \\
    35 & 4  & 175 & 175\\
    \hline
    \end{tabular}    
\end{table}

To analyse the modelling results and compare them to the observational results, we consider the time moment where the maximal observed CO depletion is reached in the model (see Table~\ref{tab:fd_age}). CO depletion factor at every point $(t,r)$ in time and space is calculated as $f_d^{(t,r)} = N^{(r)}_{\rm max}({\rm CO}) / N^{(t,r)}({\rm CO})$, where $N^{(r)}_{\rm max}({\rm CO})$ is a maximum of CO column density by time at a certain radial point $r$, and $N^{(t,r)}({\rm CO})$ is a CO column density at a point $(t,r)$.

In our modelling, the major route of $\rm H_2CO$ formation is the gas-phase reaction $\rm CH_3 + O \rightarrow H + H_2CO$ with the rate coefficient of $1.4\times 10^{-10}$~cm$^3$s$^{-1}$ inherited from the OSU database. This is the only channel for this reaction present in our default network. For core 1 it is responsible for 88\% of $\rm H_2CO$ production at the time when the observed CO depletion is achieved (the percentage for other cores is similar). Another important source of $\rm H_2CO$ is the reactive desorption in the surface reaction $\rm gH + gHCO \rightarrow gH_2CO$, which supplies 11\% of gaseous formaldehyde. The rate of $\rm H_3CO^+$ dissociative recombination resulting in the products $\rm H_2CO + H$ is three times higher than the rate of the $\rm H_2CO$ delivery to gas via chemical desorption, however, the major source of $\rm H_3CO^+$ ion is the reaction between $\rm H_2CO$ and $\rm H_3^+$. Switching off the channel $\rm H_3CO^+ + e^- \rightarrow \rm H_2CO + H$ only lowers $\rm H_2CO$ abundance by a factor of $\approx$~1.3 at the central region of the core. Thus, this reaction has a negligible impact on formaldehyde production.

In addition to the default model, we also consider a model with an additional channel for the $\rm CH_3 + O$ reaction (hereinafter the `additional channel' model). This  reaction plays an important role in combustion chemistry and was actively studied over the last four decades \citep[see Fig.~5 of][and references therein]{Xu2015}, although at elevated temperatures. Association of $\rm CH_3$ and O produces highly excited $\rm CH_3O^*$ radical which further decomposes through various accessible channels, such as: (a)~$\rm H_2CO + H$, (b)~$\rm HCO + H_2$, (c)~$\rm COH + H_2$ and (d)~$\rm CH + H_2O$ \citep{Hack2005,Xu2015}. At high temperature and high pressure conditions channels (a) and (b) dominate with the most recently estimated ratio of about 1.5. According to \cite{Xu2015}, HCO radical forming in channel (b) is unstable and can rapidly fragment into CO and H. This fragmentation is even more likely to occur at low pressure and low temperature conditions of prestellar cores in the absence of collisional cooling. The excess energy of $\rm HCO + H_2$ formation is 80.7~kcal~mol$^{-1}$ (40610~K), it is distributed among the reaction products. The reaction $\rm HCO \rightarrow CO + H$ has a barrier of 17.7~kcal~mol$^{-1}$ (8907~K). Because of low densities and low impact velocities in the interstellar medium, deexcitation rates are low too, and HCO can break down into CO and H. Initial energy excess may be also utilized to overcome the reaction barrier. With the lack of precise data on the low-temperature reaction rate and branching ratio of \ce{CH3} + O reaction, we add channel (b) as $\rm CO + H_2 + H$ for the conditions of prestellar cores to our network. We assume this to be the dominating branch due to formation of three reaction products which help to allocate the high excess energy of \ce{CH3} + O reaction. 

\begin{table}
    \caption{The channels of the $\rm CH_3 + O$ reaction and their branching ratio in our default model and in the model with additional channel.}
    \label{tab:routes}
    \centering
    \begin{tabular}{lcc}\hline
    Reaction channel & Default & Additional channel \\
    \hline
    (a) $\rm H_2CO + H$ & 1 & 1/9 \\
    (b) $\rm CO + H_2 + H$ & 0 & 8/9 \\ 
    \hline
    \end{tabular}
\end{table}

Up to our knowledge, no experimental studies report total rate of $\rm CH_3 + O$ reaction and its branching ratio at low temperatures ($\sim$10~K). In the KIDA \citep{Wakelam2012} and UMIST \citep{Millar2024} astrochemical databases, the reaction $\rm CH_3$ + $\rm O$ is present with two product channels, (a) $\rm H + H_2CO$ and (b) $\rm CO + H_2 + H$, with the branching ratio $\simeq$5:1. However, the rates are obtained in the studies of combustion or atmospheric chemistry, for the temperature ranges generally higher than that of cold cores \citep[50--2500~K;][]{Baulch1992,FockenbergPreses2002,Baulch2005,Atkinson2006,Hebrard2009}. On the other hand, some experimental and theoretical studies report temperature dependence for this reaction with the decrease of its rate constant at lower temperatures, see for example \cite{FockenbergPreses2002} and \cite{Yagi2004}. In order to describe the temperature dependence of the reaction rate, \cite{FockenbergPreses2002} suggest an Arrhenius expression for the rate coefficient $k_{\rm CH_3+O} = (2.4 \pm 0.3) \times 10^{-10}\exp(-(202 \pm 60) {\rm K}/T)$~cm$^3$~molecule$^{-1}$~s$^{-1}$ with the ``effective barrier'' equal to $202 \pm 60$~K. However, \cite{FockenbergPreses2002} stress that this expression should only be used in their investigated temperature range from 354 to 935~K. The studies of this reaction at higher temperatures ($>295$~K) report very different branching ratios \citep[see, e.g.,][]{Fockenberg1999,Preses2000,Hack2005}. Since the branching ratio is ambiguous, we investigate it as described below.

With the 1:1:0:0 branching ratio of the channels $\rm CO + H_2 + H$~:~$\rm H_2CO + H$~: $\rm COH + H_2$~: $\rm CH + H_2O$, formaldehyde abundance drops by a factor of $<1.8$ compared to the default model results. With the aim to test the impact of  temperature dependence and the limited contribution of $\rm H_2CO + H$ formation branch among the other branches of this reaction, we also performed numerical simulations with the different branching ratios between $\rm CH_3 + O \rightarrow H_2CO + H$ and $\rm CH_3 + O \rightarrow CO + H_2 + H$ reaction pathways. Factors of 2, 4, 8 and 16 were used in our simulations. The best fit to the observations is obtained with the channels ratio $\rm CO + H_2 + H : H_2CO + H = 8 : 1$ (see Table~\ref{tab:routes}), which is beyond the ratios presented in the literature before. Further increase of this ratio (e.g., to 16~:~1) has a negligible impact on $\rm H_2CO$ abundance because $\rm H_2CO$ production is further dominated by the chemical desorption in the surface reaction $\rm gH + gHCO \rightarrow gH_2CO$.

In both versions of our model, solid formaldehyde amounts to $\approx 10 \%$ of water ice within the central parts of the cores (2000 au for cores 10, 11 and 35, 4000 au for all other cores, see Fig.~\ref{fig:H2CO_CH3OH_ice}). Towards the core edge, its abundance gradually decreases to $\leq 0.1\%$ of water ice. The infrared observations reported \ce{H2CO} abundance to be 6\% of $\rm H_2O$ ice \citep{Boogert2008}.  $\rm CH_3OH$ ice follows the same radial trend with the abundance a little lower (by a factor of 1.5--4.5) than ice abundance of $\rm H_2CO$. These predictions can be used to plan infrared observations of the ices with JWST.

\subsection{Compare model and observations}

\begin{figure*}
    \centering
    \includegraphics[height=17.5cm,keepaspectratio]{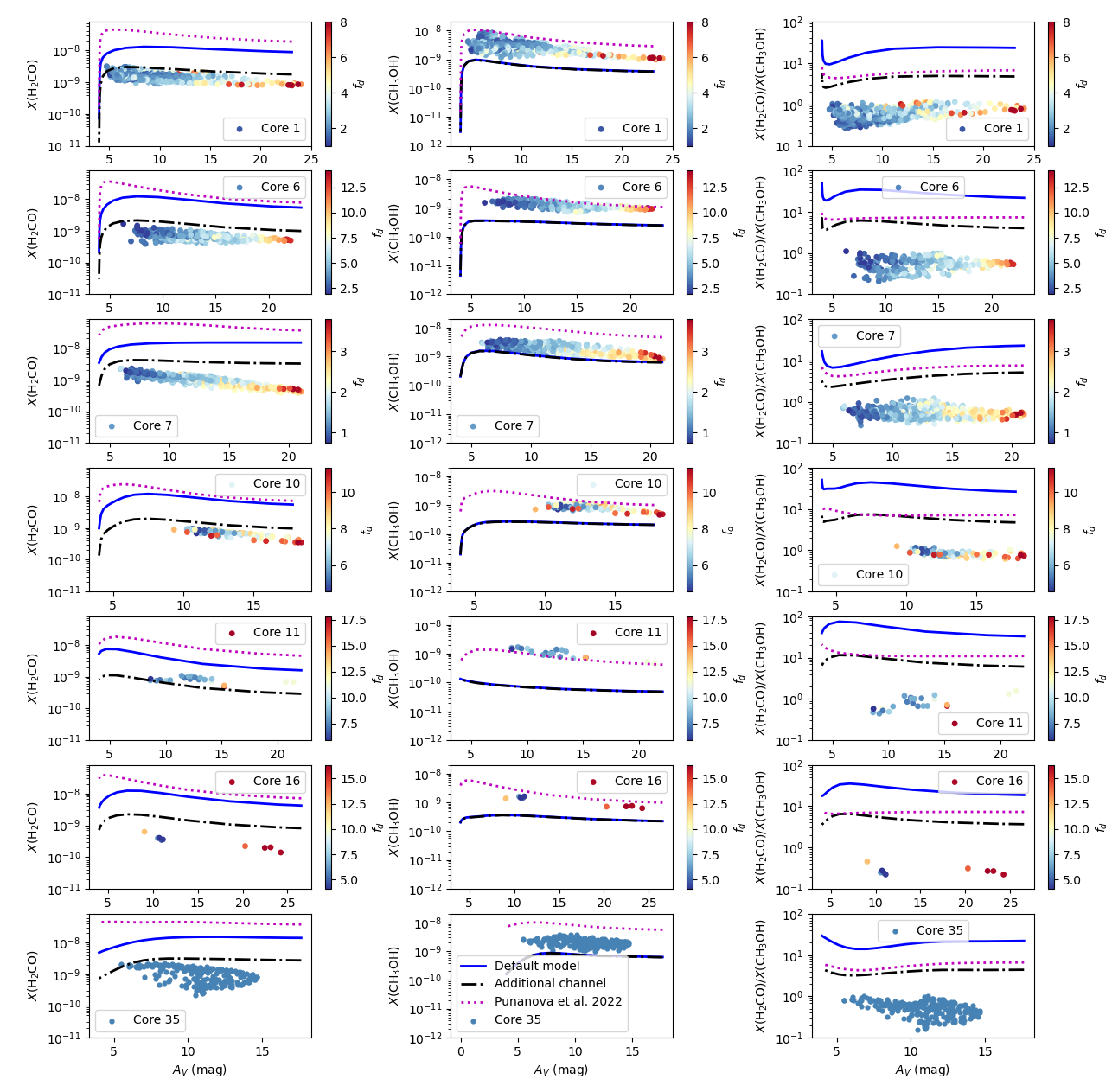}
    \caption{The comparison of the observed (coloured dots) and modelled \ce{H2CO} abundance profiles (left); CH$_3$OH abundance profiles (middle); and \ce{H2CO}:\ce{CH3OH} ratio (right) as a function of visual extinction for all cores. The colour scale represents CO depletion factor $f_d$ in the individual pixels. The data points of core 35 are all shown with the same colour since there are no CO data for this core. The models include: the default non-diffusive model (solid blue line); the non-diffusive model with the additional channel for the reaction \ce{CH3} + O (dashed-dotted black line); the old version of the model with tunnelling for diffusion of H and H$_2$ and reactive desorption efficiency from \citet{Minissale2016} as in \citet{Punanova2022} (dotted magenta line).}
    \label{fig:all-models-and-obs-meth}
\end{figure*}

In Fig.~\ref{fig:all-models-and-obs}, we compare our default model (the blue solid lines) and the model with the additional channel for the \ce{CH3} + O reaction (the dashed-dotted black lines) with observations. The left panels of Fig.~\ref{fig:all-models-and-obs} show the CO abundances. CO is an ultimate  ``sink'' for carbon in the gas phase in case of C/O ratio below 1. Thus, before freeze-out, the abundance of this species is relatively constant in dark clouds and is close to the carbon elemental abundance. The ways of CO production and destruction are well described, that is why it can serve as a first test for the models. Both models give almost the same result for CO, slightly underestimating its abundance. The small (factor of a few) underestimation may be explained by the fact that the models show only the CO from the core while in the observations we also see the CO in the surrounding cloud. Insufficient elemental abundance of carbon in the model may also caused the underestimation, as discussed in \citet{Punanova2022}.

The middle and the right panels of Fig.~\ref{fig:all-models-and-obs} show the formaldehyde abundances as functions of visual extinction and CO depletion factor. The default model overpredicts the formaldehyde abundance by an order of magnitude. However, the correlation between formaldehyde, visual extinction and CO depletion factor is reproduced well except for core 7 where both models do not show the decrease of formaldehyde abundance with $A_V$ and $f_d$. The correctly reproduced correlations with the systematic difference in the abundance may indicate that either the production rate of \ce{H2CO} is overestimated or the destruction rate of \ce{H2CO} is underestimated, while the overall model is correct. The model with the additional channel reproduces the observed formaldehyde abundance fairly well, matching the upper values (cores 1, 6, 10, 35), overestimating by a factor of a few (cores 7 and 16) or even underestimating by a factor of a few (core 11 with an embedded protostar).

With Fig.~\ref{fig:all-models-and-obs-meth}, we analyse the differences and similarities between formaldehyde and methanol, \ce{H2CO}:\ce{CH3OH} ratio, and compare the performance of the revised MONACO model with its old version. The old version of the model provides a link to the previous work where we studied methanol \citep{Punanova2022}. Just like methanol, formaldehyde decrease its abundance at higher $A_V$ and $f_d$. Unlike methanol, formaldehyde does not decrease its abundance at the core edges where $A_V$ and $f_d$ are low, even though the reliable detections of the formaldehyde lines are present at lower values of visual extinction and CO depletion factor. In fact, formaldehyde abundance keeps increasing at lower $A_V$ and $f_d$ (except for core 10 where some decrease of formaldehyde abundance at the core edge is present), which indicates the gas-phase formation route working. This detail in formaldehyde abundance distribution (better seen in Fig.~\ref{fig:all-models-and-obs}) is also missed by the chemical model, which shows a decrease in the formaldehyde abundance at lower $A_V$. 

Figure~\ref{fig:all-models-and-obs-meth} compares modelled and observed formaldehyde (left panels) and methanol (middle panels) abundances and their ratio (right panels) as functions of $A_V$. The methanol data are taken from \citet{Punanova2022}. The models include the default model (solid blue line), the model with additional channel (dashed-dotted black line), and the best model of \citet{Punanova2022} -- the diffusive model with enabled tunnelling for diffusion of hydrogen atoms and reactive desorption efficiency from \citep{Minissale2016} (hereinafter the old model, dotted magenta line). The old model overpredicts formaldehyde abundance even more than our new default model. Methanol abundance is just the same in the default model and the model with the additional channel, in both models it is underestimated by a factor of a few or matches the lower values (core 7). The old model reproduce methanol abundance better. The figure shows that the default model misses the observed relation by about an order of magnitude, while the model with the additional channel and the old model overestimate the relation by a factor of a few. Of the three models, the model with the additional channel produces the \ce{H2CO}:\ce{CH3OH} ratio, closest to observed one.

\begin{figure*}
    \centering
    \includegraphics[height=6.0cm,keepaspectratio]{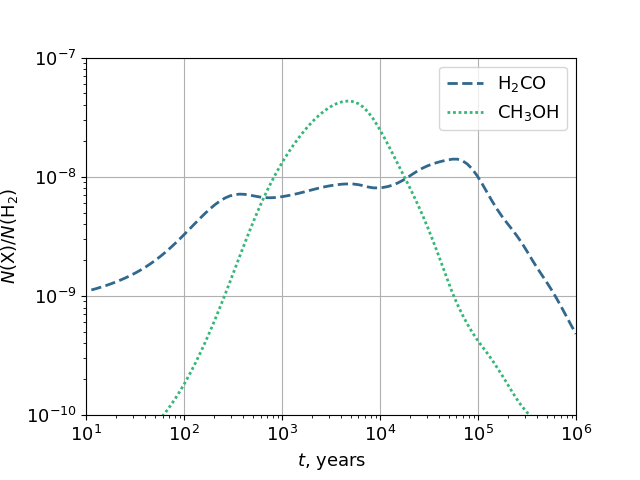}
    \includegraphics[height=6.0cm,keepaspectratio]{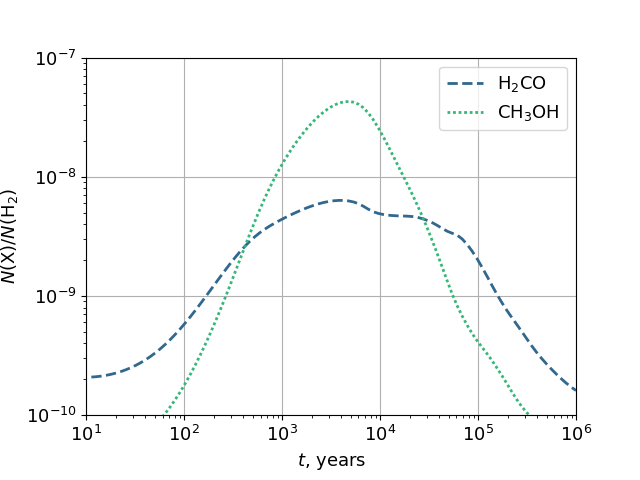}
    \caption{The time profiles of $\rm H_2CO$ and $\rm CH_3OH$ abundances derived from modelled column densities towards the centre of core 1 in our default (left) and additional channel (right) models.}
    \label{fig:H2CO_CH3OH_by_time_core01}
\end{figure*}


\section{Discussion}\label{sec:discussion}

\begin{figure*}
    \centering
    \includegraphics[height=8.5cm,keepaspectratio]{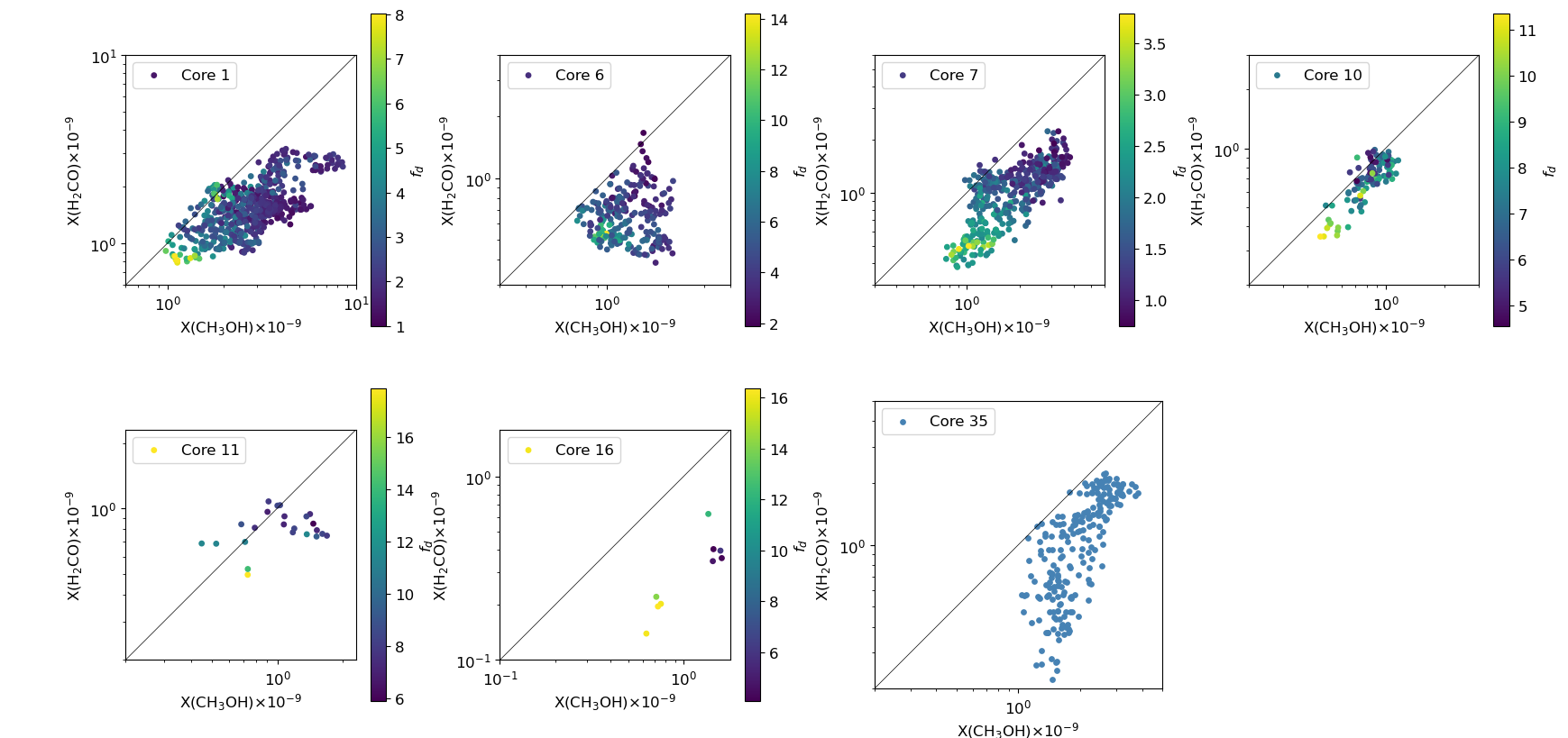}
    \caption{The abundance of formaldehyde as a function of methanol abundance. The colour scale shows CO depletion factor, $f_d$. The thin black lines show 1:1 correlation.}
    \label{fig:form-vs-meth-plot}
\end{figure*}

While in cold cores, methanol is efficiently formed only on the ice surfaces, formaldehyde can be formed both in surface and in gas phase reactions. Our model shows that the gas-phase \ce{H2CO} formation prevails, which is indirectly confirmed by the observed o/p ratio of $\simeq2$ \citep[as suggested by][]{Yocum2023}. Models predict that formaldehyde is more abundant than methanol. However, in observations, we rather see a lack of formaldehyde, which is as much or 2--5 times less abundant than methanol. We plot the methanol and formaldehyde abundances in the same coordinate scales with an indication of CO depletion factor in each position (see Fig.~\ref{fig:form-vs-meth-plot}) core by core, to analyse their correlation. This plot shows that i) formaldehyde is always less abundant then methanol; ii) both methanol and formaldehyde abundances decrease with CO depletion, starting with maximal abundances around $f_d\simeq1.5$; iii) while the correlation between the abundances looks almost linear in log-log coordinates, there is a wide spread, especially in core 6. 

\begin{figure}
    \centering
    \includegraphics[height=8.0cm,keepaspectratio]{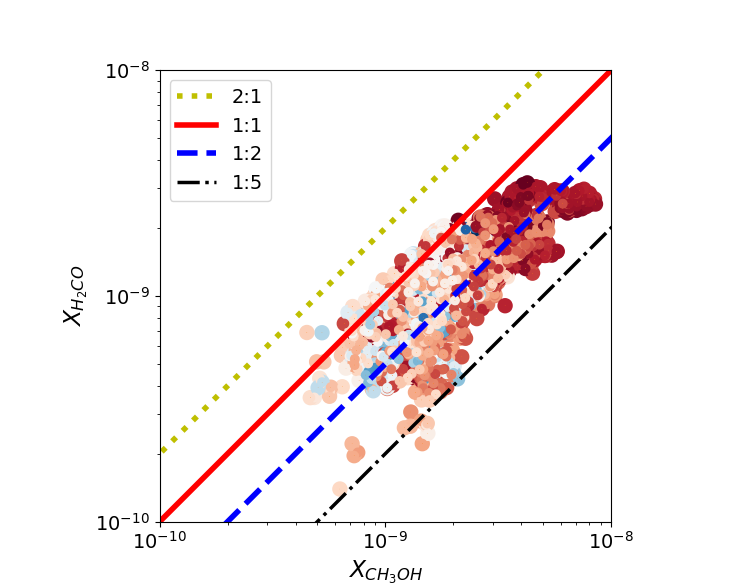}
    \caption{Abundances of methanol and formaldehyde towards the studied cores. The lines show the \ce{H2CO}:\ce{CH3OH} ratio of 1:5 (thick black dashed-dotted line), 1:2 (thick blue dashed), 1:1 correlation (thick red solid), 2:1 (thick yellow dotted). The colour represents scaled $A_V$ (no absolute values).}
    \label{fig:meth-vs-form}
\end{figure}

We tried to describe the correlation with simple linear fits and search for common trends among the regions of L1495 where the cores belong, since the regions B10, B211, B213, and B216 show different dynamic age \citep{Hacar2013,Seo2015} that might affect chemistry there, however, we did not find any common trends within the regions of the filament. Figure~\ref{fig:meth-vs-form} with all methanol and formaldehyde abundances in one plot shows that the relation between formaldehyde and methanol lie between 1:1 and 1:5. 

In \citet{Punanova2022}, tunnelling for diffusion of hydrogen atoms was essential to reproduce methanol abundance in cold cores: enabling/disabling the tunnelling impacted methanol abundance by 1--2 orders of magnitude (10$^{-11}$--10$^{-9}$ w.r.t. $n_{\rm H_2}$), while different approaches to treat reactive desorption varied methanol abundance by a factor of 2--5. The latter was not sufficient to attain the observed methanol abundances in cold cores. The new non-diffusive model exhibit the best agreement with observations when tunnelling for diffusion of hydrogen atoms is disabled.

This fact shall not be considered as a controversy between diffusive and non-diffusive rate equations-based models. It rather reflects the fact that macroscopic rate equations-based models have limited capabilities of reproducing microscopic surface effects that control hydrogen diffusion. In a series of elaborated experiments, \citet{Watanabe2010,Hama2012} and \citet{Kuwahata2015} showed that diffusion of hydrogen atoms on real surface is complex and is mainly controlled by the existing of binding sites with different energies. Tunnelling is efficient for diffusion between the low- and mid-energy binding sites. However, if a hydrogen atom is landed in a high-energy binding site that is surrounded by low- and mid-energy binding sites, it will be trapped there. Thus, at low temperatures, the diffusion of hydrogen atoms is fast but mainly short-ranged, and the answer if tunnelling is efficient for H-atoms reactivity depends on how far on average is the closest potential reactant, i.e., on surface coverage by H atoms. When translated to a macroscopic rate equations-based models, this may result in a fact that the diffusion rate of hydrogen atoms needed to explain observational results shall be higher than that caused by thermal hopping but lower than that due to quantum tunnelling. Depending on particular modelling case, the outcome is that tunnelling for diffusion of hydrogen atoms is necessary or, on the contrary, not necessary.

Our chemical simulations show that \ce{H2CO} is mainly formed in the gas phase in the reaction \ce{CH3} + O. But reactive desorption in the surface reaction gH + gHCO $\rightarrow$ g\ce{H2CO} in the methanol formation chain is also significant; at early ages (up to about $2\times 10^4$~yr), the rate of \ce{H2CO} enriching the gas via this route is comparable to the rate of the reaction \ce{CH3} + O, and sometimes even slightly higher. If the reaction \ce{CH3} + O is completely suppressed, significant amount of formaldehyde still enters the gas phase due to reactive desorption (and it is still slightly more abundant than methanol).

The non-diffusive mechanisms make a small contribution to the formation of g\ce{H2CO} itself. The diffusion reaction gH + gHCO is mainly what happens, and the rate of a similar non-diffusion reaction is 1--2 orders of magnitude lower. Non-diffusive chemistry helps to account for the COMs formation through recombination of low-mobile free radicals, e.g., recombination of CH$_{\rm n}$O radicals produced by CO + 4H $\rightarrow$ \ce{CH3OH} hydrogenation ladder \citep{Fedoseev2015,Chuang2016}. Formaldehyde is situated in the middle of this network. Thus, \ce{H2CO} mapping helps us to understand the whole picture: if there were no non-diffusive chemistry, all the surface gHCO would be spent to form g\ce{H2CO}, and not COMs because the reactions between heavy radicals proceed slowly via mechanisms of diffusive chemistry.

1D modelling of chemistry in considered cores reproduce radial distributions of observed abundances of CO, \ce{H2CO} and \ce{CH3OH} fairly well. However, as can be seen in Fig.~\ref{fig:all-models-and-obs} and~\ref{fig:all-models-and-obs-meth}, modelled abundances of those species deviate from sets of observational points. Both default and additional channel MONACO models tend to underestimate gas-phase abundances of CO and \ce{CH3OH} by a factor of 3. The reason for underestimated CO abundance can be related to the elemental abundance of carbon \citep[see][]{Punanova2022} or the fact that the model does not account for the CO in the surrounding cloud on the line of sight. In case of methanol, the key factor affecting its gas-phase abundance in cold cores is non-thermal (reactive) desorption from grains, which is poorly constrained.  However, such agreement with observations can be considered as satisfactory. Thus, detailed consideration of the impact of elemental abundances and efficiency of non-thermal desorption on the abundances of CO and \ce{CH3OH} is out of scope of this study. On the contrary, in case of \ce{H2CO}, the default MONACO model overestimated it by an order of magnitude regardless of the efficiency of reactive desorption of formaldehyde from surface. Introduction of a new branching ratio between the channels of the gas-phase reaction O + \ce{CH3} allowed us to improve the agreement between the model and observational data. This can be considered as a case of constraining routes of the chemical reaction using the astronomical observations when necessary laboratory and computational data on the reaction is missing. According to our model tests, the formaldehyde abundance is not affected by oxygen abundance, overall metallicity, abundance of C and O in the atomic form at the beginning of the core evolution or possibly higher or lower cosmic-ray ionization rate (see Sect.~\ref{Appendix} for the details).

In both default and additional channel models, methanol shows the maximum abundance $> 10^{-8}$ at $\approx 5 \times 10^3 $~yr towards the centres of the cores, which is higher than formaldehyde maximum abundance in both cases (see the example of core 1 in Figure~\ref{fig:H2CO_CH3OH_by_time_core01}; other cores have similar time patterns for these species). For core 1, $\rm CH_3OH$ abundance prevails over $\rm H_2CO$ abundance at the time interval of $7 \times 10^2$--$2 \times 10^4$~yr in our default model and at $4 \times 10^2$--$3 \times 10^4$~yr in our additional channel model. However, these intervals are located earlier than the chemical age of the core (110~kyr) calculated in agreement with the CO depletion factor.

At high CO depletion factors typical of dynamically evolved prestellar cores, CO adsorption rate significantly decreases. This results in the increase of CO hydrogenation degree by steadily adsorbing H-atoms. During this time most of the CO molecules available on the surface can be converted to the ultimate hydrogenation product, \ce{CH3OH}, rather than being partially locked in \ce{H2CO} ice. This results in the thin surface layer comprised mainly of \ce{CH3OH} ice, \citep[see, e.g.,][]{Cuppen2009,He2022}.  

The problem of the overestimated H$_2$CO abundance is not unique for the studied sources. In many chemical models formaldehyde abundance is always higher than that of methanol \citep{Walsh2009,Vasyunin2017,Chacon-Tanarro2019a,Chen2022} while many detailed observational studies report the opposite results for all (except for PDRs) kinds of star-forming regions -- low-mass cores, massive cores, IRDCs, etc. \citep[e.g.,][]{Walsh2009,Vasyunina2014,Cuadrado2017,Chacon-Tanarro2019a,Kirsanova2021}. In fact, the only sources where formaldehyde was observed to be more abundant than methanol are PDRs \citep[see][]{Guzman2013,Cuadrado2017}.

One may argue that the significant sources of ambiguity of the model results are the reaction rates. Indeed, according to the latest release of the UMIST database \citep{Millar2024}, 1448 reactions have accuracy of 25\%, 802 reactions -- of 50\%, 4631 reactions -- of factor of 2, 1854 reactions -- of an order of magnitude. The reason is that the majority of the gas phase reactions (just like the reaction \ce{CH3} + O) are investigated at room temperature while we model the medium with $T\simeq10$~K. The gas phase reactions are difficult to study in the laboratory: the particles are not charged, so it is not possible to isolate them and cool down in the ion trap by electric and/or magnetic fields. Such reactions are typically studied by cross-beams, and the free radicals are usually produced by the electron induced or photon induced dissociation which makes the produced species hot. Adiabatic expansion cools the species but cooling to 10~K is difficult. The impact of the reaction rate uncertainties on chemical modelling had been studied: due to these uncertainties the accuracy of the modelled abundances is of one order of magnitude \citep[e.g.,][]{Vasyunin2004,Vasyunin2008,Wakelam2005,Wakelam2006,Wakelam2010}. Thus, we consider the order-of-magnitude difference between the observational results and our modelling as acceptable. However, we believe that random variations of the reaction rates still do not explain the systematic underestimation of the \ce{H2CO}:\ce{CH3OH} ratio.

\section{Conclusions}\label{sec:conclusions}
In this work, we study the spatial distribution and the relative abundance of formaldehyde in seven cold dense cores of the low-mass star-forming region L1495. The obtained relative abundances of formaldehyde are compared to similar results previously obtained for methanol. We explore the correlation between formaldehyde and methanol emission and test the new astrochemical model that employs non-diffusive chemistry (Borshcheva et al. under rev. in ApJ). 
Our main findings are listed below.
\begin{enumerate}
    \item The distribution of formaldehyde in cold cores is very similar to that of methanol, with maximal abundance in the core shell, the active CO freeze-out zone, where CO depletion factor $f_d\simeq 1.5$. Like methanol, formaldehyde is depleted towards the dust peaks, with the abundance lowered by a factor of 2--10.
    \item According to our non-LTE analysis, the ortho-to-para ratio of formaldehyde is between 0.9 and 2.8 in the studied cores (2.0 on average). Following the assumption of \citet{Yocum2023} and results of our modelling, this hints for the partial gas-phase formation of formaldehyde in the cold cores.
    \item While the models predict the formaldehyde abundance by an order of magnitude higher than that of methanol ($\sim10^{-8}$ and $\sim10^{-9}$ respectively), the observed formaldehyde abundance is by a factor of 1--5 smaller than that of methanol (0.1--3.0$\times10^{-9}$ vs 0.5--8.5$\times10^{-9}$). The models though reproduce the shape of formaldehyde distribution. The reproduction of the observed profiles in the simulation runs provide a good overall validation for the applied astrochemical model. On the other hand, systematic overproduction of \ce{H2CO} in our model hints for the systematic overestimation of \ce{H2CO} formation rate or systematic underestimation of its destruction.
    \item Various tests of the upgraded model MONACO show that the most effective way to bring the modelled abundances closer to the observed ones is to introduce a new channel and apply a branching ratio to the most effective reaction that produces formaldehyde: \ce{CH3} + O $\rightarrow$ H + \ce{H2CO} \citep{Xu2015}. We put an observational constrain on the model by applying the branching ratio HCO + \ce{H2} : \ce{H2CO} + H = 8 : 1 to the products of this reaction. 
    \item The attempts to add collapse, use higher and lower cosmic-ray ionization rate, vary metallicity including sulphur and oxygen abundance, vary abundance of atomic C and O at the beginning of the core evolution, vary reaction rates within those reported in the literature, vary the gas temperature in the initial diffuse cloud did not result in a change of \ce{H2CO} : \ce{CH3OH} ratio in the model.
    \item The column density and abundance maps of formaldehyde towards the cold cores in L1495 presented in this study, along with the column density and abundance maps of methanol from \citet{Punanova2022}, can serve for testing chemical models.
\end{enumerate}

\section*{Acknowledgements}

The authors thank the anonymous referee for valuable comments that helped to improve the manuscript. Gleb Fedoseev is benefited from the Xinjiang Tianchi Talent Program (2024). Anton Vasyunin acknowledges the support of the Russian Science Foundation project 23-12-00315. This work is based on observations carried out under projects 032-14, 156-14, 013-18, 125-18 and 031-19 with the IRAM~30~m telescope. IRAM is supported by INSU/CNRS (France), MPG (Germany) and IGN (Spain).

\section*{Data Availability}

The formaldehyde spectral cubes, column density and abundance maps will be placed on the Strasbourg astronomical Data Center (CDS). The other observation data (the methanol spectral cubes, the maps of CO depletion factor, visual extinction, methanol column density, rotational temperature, abundance as well as the model abundance profiles) are published in \citet{Punanova2022}.
 



\bibliographystyle{mnras}
\bibliography{Punanova_lit}

\begin{thebibliography}{}
\makeatletter
\relax
\def\mn@urlcharsother{\let\do\@makeother \do\$\do\&\do\#\do\^\do\_\do\%\do\~}
\def\mn@doi{\begingroup\mn@urlcharsother \@ifnextchar [ {\mn@doi@}
  {\mn@doi@[]}}
\def\mn@doi@[#1]#2{\def\@tempa{#1}\ifx\@tempa\@empty \href
  {http://dx.doi.org/#2} {doi:#2}\else \href {http://dx.doi.org/#2} {#1}\fi
  \endgroup}
\def\mn@eprint#1#2{\mn@eprint@#1:#2::\@nil}
\def\mn@eprint@arXiv#1{\href {http://arxiv.org/abs/#1} {{\tt arXiv:#1}}}
\def\mn@eprint@dblp#1{\href {http://dblp.uni-trier.de/rec/bibtex/#1.xml}
  {dblp:#1}}
\def\mn@eprint@#1:#2:#3:#4\@nil{\def\@tempa {#1}\def\@tempb {#2}\def\@tempc
  {#3}\ifx \@tempc \@empty \let \@tempc \@tempb \let \@tempb \@tempa \fi \ifx
  \@tempb \@empty \def\@tempb {arXiv}\fi \@ifundefined
  {mn@eprint@\@tempb}{\@tempb:\@tempc}{\expandafter \expandafter \csname
  mn@eprint@\@tempb\endcsname \expandafter{\@tempc}}}

\bibitem[\protect\citeauthoryear{{{\'A}lvarez-Barcia}, {Russ}, {K{\"a}stner}
  \& {Lamberts}}{{{\'A}lvarez-Barcia} et~al.}{2018}]{Alvarez-Barcia2018}
{{\'A}lvarez-Barcia} S.,  {Russ} P.,  {K{\"a}stner} J.,   {Lamberts} T.,  2018,
  \mn@doi [\mnras] {10.1093/mnras/sty1478}, \href
  {https://ui.adsabs.harvard.edu/abs/2018MNRAS.479.2007A} {479, 2007}

\bibitem[\protect\citeauthoryear{{Atkinson} et~al.,}{{Atkinson}
  et~al.}{2006}]{Atkinson2006}
{Atkinson} R.,  et~al., 2006, \mn@doi [Atmospheric Chemistry \& Physics]
  {10.5194/acp-6-3625-200610.5194/acpd-5-6295-2005}, \href
  {https://ui.adsabs.harvard.edu/abs/2006ACP.....6.3625A} {6, 3625}

\bibitem[\protect\citeauthoryear{{Barnard}}{{Barnard}}{1927}]{Barnard1927}
{Barnard} E.~E.,  1927, {Catalogue of 349 dark objects in the sky}

\bibitem[\protect\citeauthoryear{{Baulch} et~al.,}{{Baulch}
  et~al.}{1992}]{Baulch1992}
{Baulch} D.~L.,  et~al., 1992, \mn@doi [Journal of Physical and Chemical
  Reference Data] {10.1063/1.555908}, \href
  {https://ui.adsabs.harvard.edu/abs/1992JPCRD..21..411B} {21, 411}

\bibitem[\protect\citeauthoryear{{Baulch} et~al.,}{{Baulch}
  et~al.}{2005}]{Baulch2005}
{Baulch} D.~L.,  et~al., 2005, \mn@doi [Journal of Physical and Chemical
  Reference Data] {10.1063/1.1748524}, \href
  {https://ui.adsabs.harvard.edu/abs/2005JPCRD..34..757B} {34, 757}

\bibitem[\protect\citeauthoryear{{Belloche}, {Garrod}, {M{\"u}ller}  \&
  {Menten}}{{Belloche} et~al.}{2014}]{Belloche2014}
{Belloche} A.,  {Garrod} R.~T.,  {M{\"u}ller} H. S.~P.,   {Menten} K.~M.,
  2014, \mn@doi [Science] {10.1126/science.1256678}, \href
  {https://ui.adsabs.harvard.edu/abs/2014Sci...345.1584B} {345, 1584}

\bibitem[\protect\citeauthoryear{{Bertin} et~al.,}{{Bertin}
  et~al.}{2016}]{Bertin2016}
{Bertin} M.,  et~al., 2016, \mn@doi [\apjl] {10.3847/2041-8205/817/2/L12},
  \href {http://adsabs.harvard.edu/abs/2016ApJ...817L..12B} {817, L12}

\bibitem[\protect\citeauthoryear{{Bocquet} et~al.,}{{Bocquet}
  et~al.}{1996}]{Bocquet1996}
{Bocquet} R.,  et~al., 1996, \mn@doi [Journal of Molecular Spectroscopy]
  {10.1006/jmsp.1996.0128}, \href
  {https://ui.adsabs.harvard.edu/abs/1996JMoSp.177..154B} {177, 154}

\bibitem[\protect\citeauthoryear{{Boogert} et~al.,}{{Boogert}
  et~al.}{2008}]{Boogert2008}
{Boogert} A.~C.~A.,  et~al., 2008, \mn@doi [\apj] {10.1086/533425}, \href
  {https://ui.adsabs.harvard.edu/abs/2008ApJ...678..985B} {678, 985}

\bibitem[\protect\citeauthoryear{{Cabedo}, {Maury}, {Girart}, {Padovani},
  {Hennebelle}, {Houde}  \& {Zhang}}{{Cabedo} et~al.}{2023}]{Cabedo2023}
{Cabedo} V.,  {Maury} A.,  {Girart} J.~M.,  {Padovani} M.,  {Hennebelle} P.,
  {Houde} M.,   {Zhang} Q.,  2023, \mn@doi [\aap]
  {10.1051/0004-6361/202243813}, \href
  {https://ui.adsabs.harvard.edu/abs/2023A&A...669A..90C} {669, A90}

\bibitem[\protect\citeauthoryear{{Chac{\'o}n-Tanarro}
  et~al.,}{{Chac{\'o}n-Tanarro} et~al.}{2019}]{Chacon-Tanarro2019a}
{Chac{\'o}n-Tanarro} A.,  et~al., 2019, \mn@doi [\aap]
  {10.1051/0004-6361/201832703}, \href
  {https://ui.adsabs.harvard.edu/abs/2019A&A...622A.141C} {622, A141}

\bibitem[\protect\citeauthoryear{{Chen}, {Chang}, {Wang}  \& {Li}}{{Chen}
  et~al.}{2022}]{Chen2022}
{Chen} L.-F.,  {Chang} Q.,  {Wang} Y.,   {Li} D.,  2022, \mn@doi [\mnras]
  {10.1093/mnras/stac2566}, \href
  {https://ui.adsabs.harvard.edu/abs/2022MNRAS.516.4627C} {516, 4627}

\bibitem[\protect\citeauthoryear{{Chuang}, {Fedoseev}, {Ioppolo}, {van
  Dishoeck}  \& {Linnartz}}{{Chuang} et~al.}{2016}]{Chuang2016}
{Chuang} K.~J.,  {Fedoseev} G.,  {Ioppolo} S.,  {van Dishoeck} E.~F.,
  {Linnartz} H.,  2016, \mn@doi [\mnras] {10.1093/mnras/stv2288}, \href
  {https://ui.adsabs.harvard.edu/abs/2016MNRAS.455.1702C} {455, 1702}

\bibitem[\protect\citeauthoryear{{Chuang}, {Fedoseev}, {Qasim}, {Ioppolo}, {van
  Dishoeck}  \& {Linnartz}}{{Chuang} et~al.}{2018}]{Chuang2018}
{Chuang} K.~J.,  {Fedoseev} G.,  {Qasim} D.,  {Ioppolo} S.,  {van Dishoeck}
  E.~F.,   {Linnartz} H.,  2018, \mn@doi [\apj] {10.3847/1538-4357/aaa24e},
  \href {https://ui.adsabs.harvard.edu/abs/2018ApJ...853..102C} {853, 102}

\bibitem[\protect\citeauthoryear{Cimas \& Largo}{Cimas \&
  Largo}{2006}]{Cimas2006}
Cimas A.,  Largo A.,  2006, \mn@doi [The Journal of Physical Chemistry A]
  {10.1021/jp0634959}, 110, 10912

\bibitem[\protect\citeauthoryear{{Cruz-Diaz}, {Mart{\'{\i}}n-Dom{\'e}nech},
  {Mu{\~n}oz Caro}  \& {Chen}}{{Cruz-Diaz} et~al.}{2016}]{Cruz-Diaz2016}
{Cruz-Diaz} G.~A.,  {Mart{\'{\i}}n-Dom{\'e}nech} R.,  {Mu{\~n}oz Caro} G.~M.,
  {Chen} Y.-J.,  2016, \mn@doi [\aap] {10.1051/0004-6361/201526761}, \href
  {http://adsabs.harvard.edu/abs/2016A%26A...592A..68C} {592, A68}

\bibitem[\protect\citeauthoryear{{Cuadrado}, {Goicoechea}, {Cernicharo},
  {Fuente}, {Pety}  \& {Tercero}}{{Cuadrado} et~al.}{2017}]{Cuadrado2017}
{Cuadrado} S.,  {Goicoechea} J.~R.,  {Cernicharo} J.,  {Fuente} A.,  {Pety} J.,
    {Tercero} B.,  2017, \mn@doi [\aap] {10.1051/0004-6361/201730459}, \href
  {https://ui.adsabs.harvard.edu/abs/2017A&A...603A.124C} {603, A124}

\bibitem[\protect\citeauthoryear{{Cuppen}, {van Dishoeck}, {Herbst}  \&
  {Tielens}}{{Cuppen} et~al.}{2009}]{Cuppen2009}
{Cuppen} H.~M.,  {van Dishoeck} E.~F.,  {Herbst} E.,   {Tielens} A.~G.~G.~M.,
  2009, \mn@doi [\aap] {10.1051/0004-6361/200913119}, \href
  {https://ui.adsabs.harvard.edu/abs/2009A&A...508..275C} {508, 275}

\bibitem[\protect\citeauthoryear{{Fayolle}, {Bertin}, {Romanzin}, {Michaut},
  {{\"O}berg}, {Linnartz}  \& {Fillion}}{{Fayolle} et~al.}{2011}]{Fayolle2011}
{Fayolle} E.~C.,  {Bertin} M.,  {Romanzin} C.,  {Michaut} X.,  {{\"O}berg}
  K.~I.,  {Linnartz} H.,   {Fillion} J.-H.,  2011, \mn@doi [\apjl]
  {10.1088/2041-8205/739/2/L36}, \href
  {https://ui.adsabs.harvard.edu/abs/2011ApJ...739L..36F} {739, L36}

\bibitem[\protect\citeauthoryear{{Fedoseev}, {Cuppen}, {Ioppolo}, {Lamberts}
  \& {Linnartz}}{{Fedoseev} et~al.}{2015}]{Fedoseev2015}
{Fedoseev} G.,  {Cuppen} H.~M.,  {Ioppolo} S.,  {Lamberts} T.,   {Linnartz} H.,
   2015, \mn@doi [\mnras] {10.1093/mnras/stu2603}, \href
  {https://ui.adsabs.harvard.edu/abs/2015MNRAS.448.1288F} {448, 1288}

\bibitem[\protect\citeauthoryear{{Fedoseev}, {Chuang}, {Ioppolo}, {Qasim}, {van
  Dishoeck}  \& {Linnartz}}{{Fedoseev} et~al.}{2017}]{Fedoseev2017}
{Fedoseev} G.,  {Chuang} K.~J.,  {Ioppolo} S.,  {Qasim} D.,  {van Dishoeck}
  E.~F.,   {Linnartz} H.,  2017, \mn@doi [\apj] {10.3847/1538-4357/aa74dc},
  \href {https://ui.adsabs.harvard.edu/abs/2017ApJ...842...52F} {842, 52}

\bibitem[\protect\citeauthoryear{{Fedoseev}, {Qasim}, {Chuang}, {Ioppolo},
  {Lamberts}, {van Dishoeck}  \& {Linnartz}}{{Fedoseev}
  et~al.}{2022}]{Fedoseev2022}
{Fedoseev} G.,  {Qasim} D.,  {Chuang} K.-J.,  {Ioppolo} S.,  {Lamberts} T.,
  {van Dishoeck} E.~F.,   {Linnartz} H.,  2022, \mn@doi [\apj]
  {10.3847/1538-4357/ac3834}, \href
  {https://ui.adsabs.harvard.edu/abs/2022ApJ...924..110F} {924, 110}

\bibitem[\protect\citeauthoryear{{Fockenberg} \& {Preses}}{{Fockenberg} \&
  {Preses}}{2002}]{FockenbergPreses2002}
{Fockenberg} C.,  {Preses} J.~M.,  2002, \mn@doi [The Journal of Physical
  Chemistry A] {10.1021/jp0141880}, 106, 2924–2930

\bibitem[\protect\citeauthoryear{{Fockenberg}, {Hall}, {Preses}, {Sears}  \&
  {Muckerman}}{{Fockenberg} et~al.}{1999}]{Fockenberg1999}
{Fockenberg} C.,  {Hall} G.~E.,  {Preses} J.~M.,  {Sears} T.~J.,   {Muckerman}
  J.~T.,  1999, \mn@doi [Journal of Physical Chemistry A] {10.1021/jp991157k},
  \href {https://ui.adsabs.harvard.edu/abs/1999JPCA..103.5722F} {103, 5722}

\bibitem[\protect\citeauthoryear{{Fontani} et~al.,}{{Fontani}
  et~al.}{2017}]{Fontani2017}
{Fontani} F.,  et~al., 2017, \mn@doi [\aap] {10.1051/0004-6361/201730527},
  \href {https://ui.adsabs.harvard.edu/abs/2017A%26A...605A..57F} {605, A57}

\bibitem[\protect\citeauthoryear{{Freeman}, {Bottinelli}, {Plume}, {Caux},
  {Monaghan}  \& {Mookerjea}}{{Freeman} et~al.}{2023}]{Freeman2023}
{Freeman} P.,  {Bottinelli} S.,  {Plume} R.,  {Caux} E.,  {Monaghan} C.,
  {Mookerjea} B.,  2023, \mn@doi [\aap] {10.1051/0004-6361/202347263}, \href
  {https://ui.adsabs.harvard.edu/abs/2023A&A...678A..18F} {678, A18}

\bibitem[\protect\citeauthoryear{{Fuchs}, {Cuppen}, {Ioppolo}, {Romanzin},
  {Bisschop}, {Andersson}, {van Dishoeck}  \& {Linnartz}}{{Fuchs}
  et~al.}{2009}]{Fuchs_ea09}
{Fuchs} G.~W.,  {Cuppen} H.~M.,  {Ioppolo} S.,  {Romanzin} C.,  {Bisschop}
  S.~E.,  {Andersson} S.,  {van Dishoeck} E.~F.,   {Linnartz} H.,  2009,
  \mn@doi [\aap] {10.1051/0004-6361/200810784}, \href
  {https://ui.adsabs.harvard.edu/abs/2009A&A...505..629F} {505, 629}

\bibitem[\protect\citeauthoryear{{Furuya}, {Aikawa}, {Hincelin}, {Hassel},
  {Bergin}, {Vasyunin}  \& {Herbst}}{{Furuya} et~al.}{2015}]{Furuya2015}
{Furuya} K.,  {Aikawa} Y.,  {Hincelin} U.,  {Hassel} G.~E.,  {Bergin} E.~A.,
  {Vasyunin} A.~I.,   {Herbst} E.,  2015, \mn@doi [\aap]
  {10.1051/0004-6361/201527050}, \href
  {https://ui.adsabs.harvard.edu/abs/2015A&A...584A.124F} {584, A124}

\bibitem[\protect\citeauthoryear{{Garrod} \& {Pauly}}{{Garrod} \&
  {Pauly}}{2011}]{GarrodPauly2011}
{Garrod} R.~T.,  {Pauly} T.,  2011, \mn@doi [\apj]
  {10.1088/0004-637X/735/1/15}, \href
  {https://ui.adsabs.harvard.edu/abs/2011ApJ...735...15G} {735, 15}

\bibitem[\protect\citeauthoryear{{Garrod}, {Wakelam}  \& {Herbst}}{{Garrod}
  et~al.}{2007}]{Garrod2007}
{Garrod} R.~T.,  {Wakelam} V.,   {Herbst} E.,  2007, \mn@doi [\aap]
  {10.1051/0004-6361:20066704}, \href
  {http://adsabs.harvard.edu/abs/2007A%26A...467.1103G} {467, 1103}

\bibitem[\protect\citeauthoryear{{Garrod}, {Jin}, {Matis}, {Jones}, {Willis}
  \& {Herbst}}{{Garrod} et~al.}{2022}]{Garrod2022}
{Garrod} R.~T.,  {Jin} M.,  {Matis} K.~A.,  {Jones} D.,  {Willis} E.~R.,
  {Herbst} E.,  2022, \mn@doi [\apjs] {10.3847/1538-4365/ac3131}, \href
  {https://ui.adsabs.harvard.edu/abs/2022ApJS..259....1G} {259, 1}

\bibitem[\protect\citeauthoryear{{Ginsburg} \& {Mirocha}}{{Ginsburg} \&
  {Mirocha}}{2011}]{Ginsburg2011}
{Ginsburg} A.,  {Mirocha} J.,  2011, {PySpecKit: Python Spectroscopic Toolkit},
  Astrophysics Source Code Library (\mn@eprint {ascl} {1109.001})

\bibitem[\protect\citeauthoryear{{Guzm{\'a}n} et~al.,}{{Guzm{\'a}n}
  et~al.}{2013}]{Guzman2013}
{Guzm{\'a}n} V.~V.,  et~al., 2013, \mn@doi [\aap]
  {10.1051/0004-6361/201322460}, \href
  {https://ui.adsabs.harvard.edu/abs/2013A&A...560A..73G} {560, A73}

\bibitem[\protect\citeauthoryear{{Hacar}, {Tafalla}, {Kauffmann}  \&
  {Kov{\'a}cs}}{{Hacar} et~al.}{2013}]{Hacar2013}
{Hacar} A.,  {Tafalla} M.,  {Kauffmann} J.,   {Kov{\'a}cs} A.,  2013, \mn@doi
  [\aap] {10.1051/0004-6361/201220090}, \href
  {http://adsabs.harvard.edu/abs/2013A%26A...554A..55H} {554, A55}

\bibitem[\protect\citeauthoryear{{Hack}, {Hold}, {Hoyermann}, {Wehmeyer}  \&
  {Zeuch}}{{Hack} et~al.}{2005}]{Hack2005}
{Hack} W.,  {Hold} M.,  {Hoyermann} K.,  {Wehmeyer} J.,   {Zeuch} T.,  2005,
  \mn@doi [Physical Chemistry Chemical Physics (Incorporating Faraday
  Transactions)] {10.1039/B419137D}, \href
  {https://ui.adsabs.harvard.edu/abs/2005PCCP....7.1977H} {7, 1977}

\bibitem[\protect\citeauthoryear{{Hama}, {Kuwahata}, {Watanabe}, {Kouchi},
  {Kimura}, {Chigai}  \& {Pirronello}}{{Hama} et~al.}{2012}]{Hama2012}
{Hama} T.,  {Kuwahata} K.,  {Watanabe} N.,  {Kouchi} A.,  {Kimura} Y.,
  {Chigai} T.,   {Pirronello} V.,  2012, \mn@doi [\apj]
  {10.1088/0004-637X/757/2/185}, \href
  {https://ui.adsabs.harvard.edu/abs/2012ApJ...757..185H} {757, 185}

\bibitem[\protect\citeauthoryear{{Harju} et~al.,}{{Harju}
  et~al.}{2017}]{Harju2017}
{Harju} J.,  et~al., 2017, \mn@doi [\aap] {10.1051/0004-6361/201628463}, \href
  {http://adsabs.harvard.edu/abs/2017A%26A...600A..61H} {600, A61}

\bibitem[\protect\citeauthoryear{{Hasegawa} \& {Herbst}}{{Hasegawa} \&
  {Herbst}}{1993}]{HasegawaHerbst93}
{Hasegawa} T.~I.,  {Herbst} E.,  1993, \mn@doi [\mnras]
  {10.1093/mnras/261.1.83}, \href
  {https://ui.adsabs.harvard.edu/abs/1993MNRAS.261...83H} {261, 83}

\bibitem[\protect\citeauthoryear{{Hasenberger} \& {Alves}}{{Hasenberger} \&
  {Alves}}{2020}]{HasenbergerAlves21}
{Hasenberger} B.,  {Alves} J.,  2020, \mn@doi [\aap]
  {10.1051/0004-6361/201936095}, \href
  {https://ui.adsabs.harvard.edu/abs/2020A&A...633A.132H} {633, A132}

\bibitem[\protect\citeauthoryear{{He} et~al.,}{{He} et~al.}{2022}]{He2022}
{He} J.,  et~al., 2022, \mn@doi [\aap] {10.1051/0004-6361/202142414}, \href
  {https://ui.adsabs.harvard.edu/abs/2022A&A...659A..65H} {659, A65}

\bibitem[\protect\citeauthoryear{{H{\'e}brard} et~al.,}{{H{\'e}brard}
  et~al.}{2009}]{Hebrard2009}
{H{\'e}brard} E.,  et~al., 2009, \mn@doi [Journal of Physical Chemistry A]
  {10.1021/jp905524e}, \href
  {https://ui.adsabs.harvard.edu/abs/2009JPCA..11311227H} {113, 11227}

\bibitem[\protect\citeauthoryear{{Herron}}{{Herron}}{1988}]{Herron1988}
{Herron} J.~T.,  1988, \mn@doi [Journal of Physical and Chemical Reference
  Data] {10.1063/1.555810}, \href
  {https://ui.adsabs.harvard.edu/abs/1988JPCRD..17..967H} {17, 967}

\bibitem[\protect\citeauthoryear{{Hidaka}, {Watanabe}, {Kouchi}  \&
  {Watanabe}}{{Hidaka} et~al.}{2009}]{Hidaka2009}
{Hidaka} H.,  {Watanabe} M.,  {Kouchi} A.,   {Watanabe} N.,  2009, \mn@doi
  [\apj] {10.1088/0004-637X/702/1/291}, \href
  {https://ui.adsabs.harvard.edu/abs/2009ApJ...702..291H} {702, 291}

\bibitem[\protect\citeauthoryear{{Hiraoka}, {Ohashi}, {Kihara}, {Yamamoto},
  {Sato}  \& {Yamashita}}{{Hiraoka} et~al.}{1994}]{Hiraoka1994}
{Hiraoka} K.,  {Ohashi} N.,  {Kihara} Y.,  {Yamamoto} K.,  {Sato} T.,
  {Yamashita} A.,  1994, \mn@doi [Chemical Physics Letters]
  {10.1016/0009-2614(94)01066-8}, \href
  {https://ui.adsabs.harvard.edu/abs/1994CPL...229..408H} {229, 408}

\bibitem[\protect\citeauthoryear{Jasper, Klippenstein, Harding  \&
  Ruscic}{Jasper et~al.}{2007}]{Jasper2007}
Jasper A.~W.,  Klippenstein S.~J.,  Harding L.~B.,   Ruscic B.,  2007, \mn@doi
  [The Journal of Physical Chemistry A] {10.1021/jp067585p}, 111, 3932

\bibitem[\protect\citeauthoryear{{Jim{\'e}nez-Serra}
  et~al.,}{{Jim{\'e}nez-Serra} et~al.}{2016}]{Jimenez-Serra2016}
{Jim{\'e}nez-Serra} I.,  et~al., 2016, \mn@doi [\apjl]
  {10.3847/2041-8205/830/1/L6}, \href
  {http://adsabs.harvard.edu/abs/2016ApJ...830L...6J} {830, L6}

\bibitem[\protect\citeauthoryear{{Jim{\'e}nez-Serra}
  et~al.,}{{Jim{\'e}nez-Serra} et~al.}{2024}]{Jimenez-Serra2024}
{Jim{\'e}nez-Serra} I.,  et~al., 2024, submitted to A\&A

\bibitem[\protect\citeauthoryear{{Jin} \& {Garrod}}{{Jin} \&
  {Garrod}}{2020}]{Jin2020}
{Jin} M.,  {Garrod} R.~T.,  2020, \mn@doi [\apjs] {10.3847/1538-4365/ab9ec8},
  \href {https://ui.adsabs.harvard.edu/abs/2020ApJS..249...26J} {249, 26}

\bibitem[\protect\citeauthoryear{{Kahane}, {Frerking}, {Langer}, {Encrenas}  \&
  {Lucas}}{{Kahane} et~al.}{1984}]{Kahane1984}
{Kahane} C.,  {Frerking} M.~A.,  {Langer} W.~D.,  {Encrenas} P.,   {Lucas} R.,
  1984, \aap, \href {https://ui.adsabs.harvard.edu/abs/1984A&A...137..211K}
  {137, 211}

\bibitem[\protect\citeauthoryear{{Kimura}, {Tsuge}, {Pirronello}, {Kouchi}  \&
  {Watanabe}}{{Kimura} et~al.}{2018}]{Kimura2018}
{Kimura} Y.,  {Tsuge} M.,  {Pirronello} V.,  {Kouchi} A.,   {Watanabe} N.,
  2018, \mn@doi [\apjl] {10.3847/2041-8213/aac102}, \href
  {https://ui.adsabs.harvard.edu/abs/2018ApJ...858L..23K} {858, L23}

\bibitem[\protect\citeauthoryear{{Kirsanova}, {Punanova}, {Semenov}  \&
  {Vasyunin}}{{Kirsanova} et~al.}{2021}]{Kirsanova2021}
{Kirsanova} M.~S.,  {Punanova} A.~F.,  {Semenov} D.~A.,   {Vasyunin} A.~I.,
  2021, \mn@doi [\mnras] {10.1093/mnras/stab2361}, \href
  {https://ui.adsabs.harvard.edu/abs/2021MNRAS.507.3810K} {507, 3810}

\bibitem[\protect\citeauthoryear{{Komichi}, {Aikawa}, {Iwasaki}  \&
  {Furuya}}{{Komichi} et~al.}{2024}]{Komichi2024}
{Komichi} Y.,  {Aikawa} Y.,  {Iwasaki} K.,   {Furuya} K.,  2024, \mn@doi
  [\mnras] {10.1093/mnras/stae2599}, \href
  {https://ui.adsabs.harvard.edu/abs/2024MNRAS.535.3738K} {535, 3738}

\bibitem[\protect\citeauthoryear{{Kuwahata}, {Hama}, {Kouchi}  \&
  {Watanabe}}{{Kuwahata} et~al.}{2015}]{Kuwahata2015}
{Kuwahata} K.,  {Hama} T.,  {Kouchi} A.,   {Watanabe} N.,  2015, \mn@doi [\prl]
  {10.1103/PhysRevLett.115.133201}, \href
  {https://ui.adsabs.harvard.edu/abs/2015PhRvL.115m3201K} {115, 133201}

\bibitem[\protect\citeauthoryear{{Lamberts}, {Fedoseev}, {van Hemert}, {Qasim},
  {Chuang}, {Santos}  \& {Linnartz}}{{Lamberts} et~al.}{2022}]{Lamberts2022}
{Lamberts} T.,  {Fedoseev} G.,  {van Hemert} M.~C.,  {Qasim} D.,  {Chuang}
  K.-J.,  {Santos} J.~C.,   {Linnartz} H.,  2022, \mn@doi [\apj]
  {10.3847/1538-4357/ac51d1}, \href
  {https://ui.adsabs.harvard.edu/abs/2022ApJ...928...48L} {928, 48}

\bibitem[\protect\citeauthoryear{{Larsson} et~al.,}{{Larsson}
  et~al.}{2007}]{Larsson2007}
{Larsson} B.,  et~al., 2007, \mn@doi [\aap] {10.1051/0004-6361:20065500}, \href
  {https://ui.adsabs.harvard.edu/abs/2007A&A...466..999L} {466, 999}

\bibitem[\protect\citeauthoryear{{Loison}, {Wakelam}  \& {Hickson}}{{Loison}
  et~al.}{2014}]{Loison2014}
{Loison} J.-C.,  {Wakelam} V.,   {Hickson} K.~M.,  2014, \mn@doi [\mnras]
  {10.1093/mnras/stu1089}, \href
  {https://ui.adsabs.harvard.edu/abs/2014MNRAS.443..398L} {443, 398}

\bibitem[\protect\citeauthoryear{{Mangum} \& {Wootten}}{{Mangum} \&
  {Wootten}}{1993}]{Mangum1993}
{Mangum} J.~G.,  {Wootten} A.,  1993, \mn@doi [\apjs] {10.1086/191841}, \href
  {https://ui.adsabs.harvard.edu/abs/1993ApJS...89..123M} {89, 123}

\bibitem[\protect\citeauthoryear{{Melnick} \& {Bergin}}{{Melnick} \&
  {Bergin}}{2005}]{Melnick2005}
{Melnick} G.~J.,  {Bergin} E.~A.,  2005, \mn@doi [Advances in Space Research]
  {10.1016/j.asr.2005.05.110}, \href
  {https://ui.adsabs.harvard.edu/abs/2005AdSpR..36.1027M} {36, 1027}

\bibitem[\protect\citeauthoryear{{Mercimek} et~al.,}{{Mercimek}
  et~al.}{2022}]{Mercimek2022}
{Mercimek} S.,  et~al., 2022, \mn@doi [\aap] {10.1051/0004-6361/202141790},
  \href {https://ui.adsabs.harvard.edu/abs/2022A&A...659A..67M} {659, A67}

\bibitem[\protect\citeauthoryear{{Millar}, {Walsh}, {Van de Sande}  \&
  {Markwick}}{{Millar} et~al.}{2024}]{Millar2024}
{Millar} T.~J.,  {Walsh} C.,  {Van de Sande} M.,   {Markwick} A.~J.,  2024,
  \mn@doi [\aap] {10.1051/0004-6361/202346908}, \href
  {https://ui.adsabs.harvard.edu/abs/2024A&A...682A.109M} {682, A109}

\bibitem[\protect\citeauthoryear{{Minissale}, {Moudens}, {Baouche}, {Chaabouni}
   \& {Dulieu}}{{Minissale} et~al.}{2016}]{Minissale2016}
{Minissale} M.,  {Moudens} A.,  {Baouche} S.,  {Chaabouni} H.,   {Dulieu} F.,
  2016, \mn@doi [\mnras] {10.1093/mnras/stw373}, \href
  {http://adsabs.harvard.edu/abs/2016MNRAS.458.2953M} {458, 2953}

\bibitem[\protect\citeauthoryear{{Minissale} et~al.,}{{Minissale}
  et~al.}{2022}]{Minissale2022}
{Minissale} M.,  et~al., 2022, \mn@doi [ACS Earth and Space Chemistry]
  {10.1021/acsearthspacechem.1c00357}, \href
  {https://ui.adsabs.harvard.edu/abs/2022ESC.....6..597M} {6, 597}

\bibitem[\protect\citeauthoryear{{Molpeceres}, {K{\"a}stner}, {Fedoseev},
  {Qasim}, {Sch{\"o}mig}, {Linnartz}  \& {Lamberts}}{{Molpeceres}
  et~al.}{2021}]{Molpeceres2021}
{Molpeceres} G.,  {K{\"a}stner} J.,  {Fedoseev} G.,  {Qasim} D.,  {Sch{\"o}mig}
  R.,  {Linnartz} H.,   {Lamberts} T.,  2021, arXiv e-prints, \href
  {https://ui.adsabs.harvard.edu/abs/2021arXiv211015887M} {p. arXiv:2110.15887}

\bibitem[\protect\citeauthoryear{{Oca{\~n}a} et~al.,}{{Oca{\~n}a}
  et~al.}{2017}]{Ocana2017}
{Oca{\~n}a} A.~J.,  et~al., 2017, \mn@doi [\apj] {10.3847/1538-4357/aa93d9},
  \href {https://ui.adsabs.harvard.edu/abs/2017ApJ...850...28O} {850, 28}

\bibitem[\protect\citeauthoryear{{Olofsson}}{{Olofsson}}{2006}]{Olofsson2006}
{Olofsson} H.,  2006, in Complex Molecules in Space: Present Status and
  Prospects with ALMA. p.~6

\bibitem[\protect\citeauthoryear{{Padovani}, {Marcowith}, {Hennebelle}  \&
  {Ferri{\`e}re}}{{Padovani} et~al.}{2016}]{Padovani2016}
{Padovani} M.,  {Marcowith} A.,  {Hennebelle} P.,   {Ferri{\`e}re} K.,  2016,
  \mn@doi [\aap] {10.1051/0004-6361/201628221}, \href
  {https://ui.adsabs.harvard.edu/abs/2016A&A...590A...8P} {590, A8}

\bibitem[\protect\citeauthoryear{{Pagani} et~al.,}{{Pagani}
  et~al.}{2003}]{Pagani2003}
{Pagani} L.,  et~al., 2003, \mn@doi [\aap] {10.1051/0004-6361:20030344}, \href
  {https://ui.adsabs.harvard.edu/abs/2003A&A...402L..77P} {402, L77}

\bibitem[\protect\citeauthoryear{{Palmeirim} et~al.,}{{Palmeirim}
  et~al.}{2013}]{Palmeirim2013}
{Palmeirim} P.,  et~al., 2013, \mn@doi [\aap] {10.1051/0004-6361/201220500},
  \href {http://adsabs.harvard.edu/abs/2013A%26A...550A..38P} {550, A38}

\bibitem[\protect\citeauthoryear{{Pickett}, {Poynter}, {Cohen}, {Delitsky},
  {Pearson}  \& {M{\"u}ller}}{{Pickett} et~al.}{1998}]{Pickett1998}
{Pickett} H.~M.,  {Poynter} R.~L.,  {Cohen} E.~A.,  {Delitsky} M.~L.,
  {Pearson} J.~C.,   {M{\"u}ller} H.~S.~P.,  1998, \mn@doi [\jqsrt]
  {10.1016/S0022-4073(98)00091-0}, \href
  {http://adsabs.harvard.edu/abs/1998JQSRT..60..883P} {60, 883}

\bibitem[\protect\citeauthoryear{{Potapov} \& {Garrod}}{{Potapov} \&
  {Garrod}}{2024}]{Potapov2024}
{Potapov} A.,  {Garrod} R.~T.,  2024, \mn@doi [\aap]
  {10.1051/0004-6361/202450958}, \href
  {https://ui.adsabs.harvard.edu/abs/2024A&A...692A.252P} {692, A252}

\bibitem[\protect\citeauthoryear{{Prasad} \& {Tarafdar}}{{Prasad} \&
  {Tarafdar}}{1983}]{PrasadTarafdar1983}
{Prasad} S.~S.,  {Tarafdar} S.~P.,  1983, \mn@doi [\apj] {10.1086/160896},
  \href {https://ui.adsabs.harvard.edu/abs/1983ApJ...267..603P} {267, 603}

\bibitem[\protect\citeauthoryear{{Preses}, {Fockenberg}  \& {Flynn}}{{Preses}
  et~al.}{2000}]{Preses2000}
{Preses} J.~M.,  {Fockenberg} C.,   {Flynn} G.~W.,  2000, \mn@doi [Journal of
  Physical Chemistry A] {10.1021/jp000404d}, \href
  {https://ui.adsabs.harvard.edu/abs/2000JPCA..104.6758P} {104, 6758}

\bibitem[\protect\citeauthoryear{{Punanova}, {Caselli}, {Pineda}, {Pon},
  {Tafalla}, {Hacar}  \& {Bizzocchi}}{{Punanova} et~al.}{2018}]{Punanova2018b}
{Punanova} A.,  {Caselli} P.,  {Pineda} J.~E.,  {Pon} A.,  {Tafalla} M.,
  {Hacar} A.,   {Bizzocchi} L.,  2018, \mn@doi [\aap]
  {10.1051/0004-6361/201731159}, \href
  {http://adsabs.harvard.edu/abs/2018A%26A...617A..27P} {617, A27}

\bibitem[\protect\citeauthoryear{{Punanova}, {Vasyunin}, {Caselli}, {Howard},
  {Spezzano}, {Shirley}, {Scibelli}  \& {Harju}}{{Punanova}
  et~al.}{2022}]{Punanova2022}
{Punanova} A.,  {Vasyunin} A.,  {Caselli} P.,  {Howard} A.,  {Spezzano} S.,
  {Shirley} Y.,  {Scibelli} S.,   {Harju} J.,  2022, \mn@doi [\apj]
  {10.3847/1538-4357/ac4e7d}, \href
  {https://ui.adsabs.harvard.edu/abs/2022ApJ...927..213P} {927, 213}

\bibitem[\protect\citeauthoryear{{Rawlings}, {Hartquist}, {Menten}  \&
  {Williams}}{{Rawlings} et~al.}{1992}]{Rawlings1992}
{Rawlings} J.~M.~C.,  {Hartquist} T.~W.,  {Menten} K.~M.,   {Williams} D.~A.,
  1992, \mn@doi [\mnras] {10.1093/mnras/255.3.471}, \href
  {https://ui.adsabs.harvard.edu/abs/1992MNRAS.255..471R} {255, 471}

\bibitem[\protect\citeauthoryear{{Roccatagliata}, {Franciosini}, {Sacco}, {Rand
  ich}  \& {Sicilia-Aguilar}}{{Roccatagliata} et~al.}{2020}]{Roccatagliata2020}
{Roccatagliata} V.,  {Franciosini} E.,  {Sacco} G.~G.,  {Rand ich} S.,
  {Sicilia-Aguilar} A.,  2020, \mn@doi [\aap] {10.1051/0004-6361/201936401},
  \href {https://ui.adsabs.harvard.edu/abs/2020A&A...638A..85R} {638, A85}

\bibitem[\protect\citeauthoryear{{Ruaud}, {Loison}, {Hickson}, {Gratier},
  {Hersant}  \& {Wakelam}}{{Ruaud} et~al.}{2015}]{Ruaud2015}
{Ruaud} M.,  {Loison} J.~C.,  {Hickson} K.~M.,  {Gratier} P.,  {Hersant} F.,
  {Wakelam} V.,  2015, \mn@doi [\mnras] {10.1093/mnras/stu2709}, \href
  {https://ui.adsabs.harvard.edu/abs/2015MNRAS.447.4004R} {447, 4004}

\bibitem[\protect\citeauthoryear{{Santiago-Garc{\'{\i}}a}, {Tafalla},
  {Johnstone}  \& {Bachiller}}{{Santiago-Garc{\'{\i}}a}
  et~al.}{2009}]{Santiago-Garcia2009}
{Santiago-Garc{\'{\i}}a} J.,  {Tafalla} M.,  {Johnstone} D.,   {Bachiller} R.,
  2009, \mn@doi [\aap] {10.1051/0004-6361:200810739}, \href
  {http://adsabs.harvard.edu/abs/2009A%26A...495..169S} {495, 169}

\bibitem[\protect\citeauthoryear{{Santos}, {Chuang}, {Lamberts}, {Fedoseev},
  {Ioppolo}  \& {Linnartz}}{{Santos} et~al.}{2022}]{Santos2022}
{Santos} J.~C.,  {Chuang} K.-J.,  {Lamberts} T.,  {Fedoseev} G.,  {Ioppolo} S.,
    {Linnartz} H.,  2022, \mn@doi [\apjl] {10.3847/2041-8213/ac7158}, \href
  {https://ui.adsabs.harvard.edu/abs/2022ApJ...931L..33S} {931, L33}

\bibitem[\protect\citeauthoryear{{Schlafly} et~al.,}{{Schlafly}
  et~al.}{2014}]{Schlafly2014}
{Schlafly} E.~F.,  et~al., 2014, \mn@doi [\apj] {10.1088/0004-637X/786/1/29},
  \href {http://adsabs.harvard.edu/abs/2014ApJ...786...29S} {786, 29}

\bibitem[\protect\citeauthoryear{{Sch{\"o}ier}, {van der Tak}, {van Dishoeck}
  \& {Black}}{{Sch{\"o}ier} et~al.}{2005}]{Schoeier2005}
{Sch{\"o}ier} F.~L.,  {van der Tak} F.~F.~S.,  {van Dishoeck} E.~F.,   {Black}
  J.~H.,  2005, \mn@doi [\aap] {10.1051/0004-6361:20041729}, \href
  {http://esoads.eso.org/abs/2005A%26A...432..369S} {432, 369}

\bibitem[\protect\citeauthoryear{{Scibelli} \& {Shirley}}{{Scibelli} \&
  {Shirley}}{2020}]{Scibelli2020}
{Scibelli} S.,  {Shirley} Y.,  2020, \mn@doi [\apj] {10.3847/1538-4357/ab7375},
  \href {https://ui.adsabs.harvard.edu/abs/2020ApJ...891...73S} {891, 73}

\bibitem[\protect\citeauthoryear{{Scibelli}, {Shirley}, {Vasyunin}  \&
  {Launhardt}}{{Scibelli} et~al.}{2021}]{Scibelli2021}
{Scibelli} S.,  {Shirley} Y.,  {Vasyunin} A.,   {Launhardt} R.,  2021, \mn@doi
  [\mnras] {10.1093/mnras/stab1151}, \href
  {https://ui.adsabs.harvard.edu/abs/2021MNRAS.504.5754S} {504, 5754}

\bibitem[\protect\citeauthoryear{{Seo} et~al.,}{{Seo} et~al.}{2015}]{Seo2015}
{Seo} Y.~M.,  et~al., 2015, \mn@doi [\apj] {10.1088/0004-637X/805/2/185}, \href
  {http://adsabs.harvard.edu/abs/2015ApJ...805..185S} {805, 185}

\bibitem[\protect\citeauthoryear{{Sipil{\"a}}, {Caselli}  \&
  {Harju}}{{Sipil{\"a}} et~al.}{2013}]{Sipila2013}
{Sipil{\"a}} O.,  {Caselli} P.,   {Harju} J.,  2013, \mn@doi [\aap]
  {10.1051/0004-6361/201220922}, \href
  {http://adsabs.harvard.edu/abs/2013A%26A...554A..92S} {554, A92}

\bibitem[\protect\citeauthoryear{{Sipil{\"a}}, {Zhao}  \&
  {Caselli}}{{Sipil{\"a}} et~al.}{2020}]{Sipila2020}
{Sipil{\"a}} O.,  {Zhao} B.,   {Caselli} P.,  2020, \mn@doi [\aap]
  {10.1051/0004-6361/202038353}, \href
  {https://ui.adsabs.harvard.edu/abs/2020A&A...640A..94S} {640, A94}

\bibitem[\protect\citeauthoryear{{Tafalla} \& {Hacar}}{{Tafalla} \&
  {Hacar}}{2015}]{Tafalla2015}
{Tafalla} M.,  {Hacar} A.,  2015, \mn@doi [\aap] {10.1051/0004-6361/201424576},
  \href {http://adsabs.harvard.edu/abs/2015A%26A...574A.104T} {574, A104}

\bibitem[\protect\citeauthoryear{{Taquet}, {Wirstr{\"o}m}, {Charnley}, {Faure},
  {L{\'o}pez-Sepulcre}  \& {Persson}}{{Taquet} et~al.}{2017}]{Taquet2017}
{Taquet} V.,  {Wirstr{\"o}m} E.~S.,  {Charnley} S.~B.,  {Faure} A.,
  {L{\'o}pez-Sepulcre} A.,   {Persson} C.~M.,  2017, \mn@doi [\aap]
  {10.1051/0004-6361/201630023}, \href
  {https://ui.adsabs.harvard.edu/abs/2017A&A...607A..20T} {607, A20}

\bibitem[\protect\citeauthoryear{{Vastel}, {Ceccarelli}, {Lefloch}  \&
  {Bachiller}}{{Vastel} et~al.}{2014}]{Vastel2014}
{Vastel} C.,  {Ceccarelli} C.,  {Lefloch} B.,   {Bachiller} R.,  2014, \mn@doi
  [\apjl] {10.1088/2041-8205/795/1/L2}, \href
  {http://adsabs.harvard.edu/abs/2014ApJ...795L...2V} {795, L2}

\bibitem[\protect\citeauthoryear{{Vasyunin} \& {Herbst}}{{Vasyunin} \&
  {Herbst}}{2013}]{Vasyunin2013}
{Vasyunin} A.~I.,  {Herbst} E.,  2013, \mn@doi [\apj]
  {10.1088/0004-637X/769/1/34}, \href
  {http://adsabs.harvard.edu/abs/2013ApJ...769...34V} {769, 34}

\bibitem[\protect\citeauthoryear{{Vasyunin}, {Sobolev}, {Wiebe}  \&
  {Semenov}}{{Vasyunin} et~al.}{2004}]{Vasyunin2004}
{Vasyunin} A.~I.,  {Sobolev} A.~M.,  {Wiebe} D.~S.,   {Semenov} D.~A.,  2004,
  \mn@doi [Astronomy Letters] {10.1134/1.1784498}, \href
  {http://adsabs.harvard.edu/abs/2004AstL...30..566V} {30, 566}

\bibitem[\protect\citeauthoryear{{Vasyunin}, {Semenov}, {Henning}, {Wakelam},
  {Herbst}  \& {Sobolev}}{{Vasyunin} et~al.}{2008}]{Vasyunin2008}
{Vasyunin} A.~I.,  {Semenov} D.,  {Henning} T.,  {Wakelam} V.,  {Herbst} E.,
  {Sobolev} A.~M.,  2008, \mn@doi [\apj] {10.1086/523887}, \href
  {http://adsabs.harvard.edu/abs/2008ApJ...672..629V} {672, 629}

\bibitem[\protect\citeauthoryear{{Vasyunin}, {Caselli}, {Dulieu}  \&
  {Jim{\'e}nez-Serra}}{{Vasyunin} et~al.}{2017}]{Vasyunin2017}
{Vasyunin} A.~I.,  {Caselli} P.,  {Dulieu} F.,   {Jim{\'e}nez-Serra} I.,  2017,
  \mn@doi [\apj] {10.3847/1538-4357/aa72ec}, \href
  {http://adsabs.harvard.edu/abs/2017ApJ...842...33V} {842, 33}

\bibitem[\protect\citeauthoryear{{Vasyunina}, {Vasyunin}, {Herbst}, {Linz},
  {Voronkov}, {Britton}, {Zinchenko}  \& {Schuller}}{{Vasyunina}
  et~al.}{2014}]{Vasyunina2014}
{Vasyunina} T.,  {Vasyunin} A.~I.,  {Herbst} E.,  {Linz} H.,  {Voronkov} M.,
  {Britton} T.,  {Zinchenko} I.,   {Schuller} F.,  2014, \mn@doi [\apj]
  {10.1088/0004-637X/780/1/85}, \href
  {https://ui.adsabs.harvard.edu/abs/2014ApJ...780...85V} {780, 85}

\bibitem[\protect\citeauthoryear{{Wakelam} \& {Herbst}}{{Wakelam} \&
  {Herbst}}{2008}]{Wakelam2008}
{Wakelam} V.,  {Herbst} E.,  2008, \mn@doi [\apj] {10.1086/587734}, \href
  {https://ui.adsabs.harvard.edu/abs/2008ApJ...680..371W} {680, 371}

\bibitem[\protect\citeauthoryear{{Wakelam}, {Selsis}, {Herbst}  \&
  {Caselli}}{{Wakelam} et~al.}{2005}]{Wakelam2005}
{Wakelam} V.,  {Selsis} F.,  {Herbst} E.,   {Caselli} P.,  2005, \mn@doi [\aap]
  {10.1051/0004-6361:20053673}, \href
  {https://ui.adsabs.harvard.edu/abs/2005A&A...444..883W} {444, 883}

\bibitem[\protect\citeauthoryear{{Wakelam}, {Herbst}  \& {Selsis}}{{Wakelam}
  et~al.}{2006}]{Wakelam2006}
{Wakelam} V.,  {Herbst} E.,   {Selsis} F.,  2006, \mn@doi [\aap]
  {10.1051/0004-6361:20054682}, \href
  {https://ui.adsabs.harvard.edu/abs/2006A&A...451..551W} {451, 551}

\bibitem[\protect\citeauthoryear{{Wakelam} et~al.,}{{Wakelam}
  et~al.}{2010}]{Wakelam2010}
{Wakelam} V.,  et~al., 2010, \mn@doi [\ssr] {10.1007/s11214-010-9712-5}, \href
  {https://ui.adsabs.harvard.edu/abs/2010SSRv..156...13W} {156, 13}

\bibitem[\protect\citeauthoryear{{Wakelam} et~al.,}{{Wakelam}
  et~al.}{2012}]{Wakelam2012}
{Wakelam} V.,  et~al., 2012, \mn@doi [\apjs] {10.1088/0067-0049/199/1/21},
  \href {https://ui.adsabs.harvard.edu/abs/2012ApJS..199...21W} {199, 21}

\bibitem[\protect\citeauthoryear{{Walsh}, {Harada}, {Herbst}  \&
  {Millar}}{{Walsh} et~al.}{2009}]{Walsh2009}
{Walsh} C.,  {Harada} N.,  {Herbst} E.,   {Millar} T.~J.,  2009, \mn@doi [\apj]
  {10.1088/0004-637X/700/1/752}, \href
  {https://ui.adsabs.harvard.edu/abs/2009ApJ...700..752W} {700, 752}

\bibitem[\protect\citeauthoryear{{Watanabe} \& {Kouchi}}{{Watanabe} \&
  {Kouchi}}{2002}]{Watanabe2002}
{Watanabe} N.,  {Kouchi} A.,  2002, \mn@doi [\apjl] {10.1086/341412}, \href
  {http://adsabs.harvard.edu/abs/2002ApJ...571L.173W} {571, L173}

\bibitem[\protect\citeauthoryear{{Watanabe}, {Kimura}, {Kouchi}, {Chigai},
  {Hama}  \& {Pirronello}}{{Watanabe} et~al.}{2010}]{Watanabe2010}
{Watanabe} N.,  {Kimura} Y.,  {Kouchi} A.,  {Chigai} T.,  {Hama} T.,
  {Pirronello} V.,  2010, \mn@doi [\apjl] {10.1088/2041-8205/714/2/L233}, \href
  {https://ui.adsabs.harvard.edu/abs/2010ApJ...714L.233W} {714, L233}

\bibitem[\protect\citeauthoryear{{Wiesenfeld} \& {Faure}}{{Wiesenfeld} \&
  {Faure}}{2013}]{Wisenfeld2013}
{Wiesenfeld} L.,  {Faure} A.,  2013, \mn@doi [\mnras] {10.1093/mnras/stt616},
  \href {https://ui.adsabs.harvard.edu/abs/2013MNRAS.432.2573W} {432, 2573}

\bibitem[\protect\citeauthoryear{{Xu}, {Raghunath}  \& {Lin}}{{Xu}
  et~al.}{2015}]{Xu2015}
{Xu} Z.~F.,  {Raghunath} P.,   {Lin} M.~C.,  2015, \mn@doi [Journal of Physical
  Chemistry A] {10.1021/acs.jpca.5b00553}, \href
  {https://ui.adsabs.harvard.edu/abs/2015JPCA..119.7404X} {119, 7404}

\bibitem[\protect\citeauthoryear{{Yagi}, {Takayanagi}, {Taketsugu}  \&
  {Hirao}}{{Yagi} et~al.}{2004}]{Yagi2004}
{Yagi} K.,  {Takayanagi} T.,  {Taketsugu} T.,   {Hirao} K.,  2004, \mn@doi
  [\jcp] {10.1063/1.1737732}, \href
  {https://ui.adsabs.harvard.edu/abs/2004JChPh.12010395Y} {120, 10395}

\bibitem[\protect\citeauthoryear{{Yang} et~al.,}{{Yang}
  et~al.}{2022}]{Yang2022}
{Yang} Y.-L.,  et~al., 2022, \mn@doi [\apjl] {10.3847/2041-8213/aca289}, \href
  {https://ui.adsabs.harvard.edu/abs/2022ApJ...941L..13Y} {941, L13}

\bibitem[\protect\citeauthoryear{{Yocum}, {Wilkins}, {Bardwell}, {Milam}  \&
  {Gerakines}}{{Yocum} et~al.}{2023}]{Yocum2023}
{Yocum} K.~M.,  {Wilkins} O.~H.,  {Bardwell} J.~C.,  {Milam} S.~N.,
  {Gerakines} P.~A.,  2023, \mn@doi [\apjl] {10.3847/2041-8213/ad0bee}, \href
  {https://ui.adsabs.harvard.edu/abs/2023ApJ...958L..41Y} {958, L41}

\bibitem[\protect\citeauthoryear{Yu, Yang  \& Lin}{Yu et~al.}{1993}]{Yu1993}
Yu T.,  Yang D.~L.,   Lin M.~C.,  1993, \mn@doi [International Journal of
  Chemical Kinetics] {https://doi.org/10.1002/kin.550251210}, 25, 1053

\bibitem[\protect\citeauthoryear{de Souza~Machado, Martins, Baptista  \&
  Bauerfeldt}{de~Souza~Machado et~al.}{2020}]{Machado2020}
de Souza~Machado G.,  Martins E.~M.,  Baptista L.,   Bauerfeldt G.~F.,  2020,
  \mn@doi [The Journal of Physical Chemistry A] {10.1021/acs.jpca.9b11690},
  124, 2309

\bibitem[\protect\citeauthoryear{{van der Tak}, {Black}, {Sch{\"o}ier},
  {Jansen}  \& {van Dishoeck}}{{van der Tak} et~al.}{2007}]{vanderTak2007}
{van der Tak} F.~F.~S.,  {Black} J.~H.,  {Sch{\"o}ier} F.~L.,  {Jansen} D.~J.,
   {van Dishoeck} E.~F.,  2007, \mn@doi [\aap] {10.1051/0004-6361:20066820},
  \href {http://adsabs.harvard.edu/abs/2007A%26A...468..627V} {468, 627}

\makeatother
\end{thebibliography}



\appendix

\section{Other model tests}\label{Appendix}

In this section, we present our attempts to modify the model to see the impact on the formaldehyde-to-methanol ratio, those did not bring a sufficient result. We varied the cosmic-ray ionization rate, the reaction rates, metallicity and the oxygen abundance, abundance of atomic C and O at the beginning of the core evolution, introduced gravitational collapse and added several reactions to our network.

\subsection{Cosmic-ray ionization rate} 

Cosmic-ray ionization rate is known to vary in star forming regions, possibly because of the acceleration of energetic particles in outflows along young stellar objects \citep[e.g.,][]{Padovani2016,Fontani2017,Cabedo2023}.  With the default model, we varied the cosmic-ray ionization rate: $\zeta=1.3\cdot10^{-16}$~s$^{-1}$, $1.3\cdot10^{-17}$~s$^{-1}$, $1.3\cdot10^{-18}$~s$^{-1}$. With $\zeta=1.3\cdot10^{-16}$~s$^{-1}$, at the age corresponding to the observed CO depletion, obtained $\rm CH_3OH$ and $\rm H_2CO$ abundances differ by a factor of $\leq 1.5$ from their default abundances. With $\zeta=1.3\cdot10^{-18}$~s$^{-1}$, $\rm CH_3OH$ abundance drops by a factor of 2.0 in the core centre and does not change in the outer parts of the core, and $\rm H_2CO$ abundance drops by a factor of $\leq 3 $ over all the radii. Thus, the ratio between $\rm H_2CO$ and $\rm CH_3OH$ abundances still remains high.

\subsection{Varying oxygen abundance and metallicity} 

Some of the latest studies \citep[e.g.,][]{Harju2017,Scibelli2021} suggest that chemical modelling of cold cores in the local ISM may require higher initial elemental abundances of nitrogen or sulphur than the standard ``low-metal'' set \citep[EA1,][]{Wakelam2008}. At the same time, the attempts to detect \ce{O2} with the ODIN satellite towards the star-forming regions \citep{Pagani2003,Melnick2005,Olofsson2006,Larsson2007} showed that the oxygen abundance may be much lower than expected (X(\ce{O2}) wrt \ce{H2} = 5$\cdot10^{-8}$). 

We varied the atomic oxygen abundance from $5 \times 10^{-4}$ to $1\times10^{-4}$ with a step of $1\times10^{-4}$ and from $9 \times 10^{-5}$ to $5 \times 10^{-5}$ with a step of $1\times10^{-5}$ (in EA1, $X$(O)=1.76$\times10^{-4}$), with the other initial elemental abundances corresponding to the EA1 ``low-metal'' of \citet{Wakelam2008}. However, high or low initial oxygen abundance does not decrease $\rm H_2CO$~:~$\rm CH_3OH$ abundance ratio. At the core centre, the highest $\rm H_2CO$ abundance is $1.1 \times 10^{-8}$ when the oxygen abundance is $9 \times 10^{-5}$. Increase or decrease in oxygen abundance within the aforementioned range lowers $\rm H_2CO$ abundance by a factor of $\leq 1.5$. As for $\rm CH_3OH$, at the core centre, its highest abundance equals $3.5 \times 10^{-10}$ for the oxygen abundance of $2.0 \times 10^{-4}$. Increase or decrease in oxygen abundance within the aforementioned range lowers $\rm CH_3OH$ abundance by a factor of $\leq 1.1$ at the core centre and by more than an order of magnitude at the edge of the core.

If we apply ``high-metal'' abundances \citep[EA2,][]{Wakelam2008}, at the core centre, $\rm H_2CO$ abundance drops to $3.2 \times 10^{-9}$, but also $\rm CH_3OH$ abundance drops and comes to $4.1 \times 10^{-11}$. Varying initial oxygen abundance (and keeping other initial abundances as in the ``high-metal'' case) does not decrease $\rm H_2CO$~:~$\rm CH_3OH$ abundance ratio. 

\subsection{The model with collapse}

The static models are limited and sometimes modelling of the cloud collapse is needed to adequately reproduce chemical processes. We included the collapse to our model, as it is defined in \citet{Rawlings1992}. In addition to the free-fall collapse ($B = 1$), we launched fast ($B = 5$, 10, 100) and slow ($B = 0.3$) collapse. We also launched collapse with a decrease in temperature, as was done in \citet{GarrodPauly2011} and in \citet{Jimenez-Serra2024}. In this case, at the stage of a translucent cloud, the temperature remains constant (15~K).

Free-fall collapse ($B = 1$) and the collapse with $B = 0.3$ result in a reverse radial profile of $\rm CH_3OH$ -- its abundance drops towards the edge of the core which contradicts the observations. The models with a higher $B$ do not change the profile of $\rm CH_3OH$ abundance. None of the models with collapse produce less $\rm H_2CO$ than $\rm CH_3OH$ at the moment of the observed CO depletion.

One of the aims for performing collapse is testing our assumption that at the chemical ages of prestellar cores typical for static models ($\sim$~$10^5$~years), atomic oxygen is not yet significantly incorporated in other species. However, in the models with slow collapse, its abundance does not show a significant drop until the last stage of the collapse. While CO is not frozen out, O is efficiently replenished by the reaction CO + \ce{He^+} $\rightarrow$ \ce{C^+} + O + He. As for molecular oxygen, until the last stage of the collapse, its abundance gradually increases mainly due to the reaction OH + O $\rightarrow$ \ce{O2} + H. After that, when gas density becomes high enough, \ce{O2} abundance drops rapidly because of strong freeze-out. However, at the time when the observed CO depletion is reached (see Table~\ref{tab:fd_age}), \ce{O2} abundance is still higher than $10^{-7}$ in the most part of the core (within $R< 8000$~au), which is not in agreement with the observations for similar objects. The effective way to reduce the modelled \ce{O2} abundance is a moderate reduction of the initial abundance of atomic oxygen. However, if the initial O abundance is reduced to $<10^{-4}$, the $\rm CH_3OH$ profile becomes reverse (with high abundance towards the centre and low abundance towards the edges) for all tested collapse models.

\subsection{Reducing C and O abundance at the start of the core stage}

Before running prestellar core models, we consider the chemical evolution of the translucent cloud stage. At the final time moment of this stage, free carbon abundance equals $2.7 \cdot 10^{-6}$ and free oxygen abundance is $8.5 \cdot 10^{-5}$. Since formaldehyde is mainly produced in the reaction $\rm CH_3 + O \rightarrow H_2CO + H$, the lowering of atomic C and especially O abundances could suppress its formation. In a recent modelling study by \cite{Komichi2024}, who investigate the chemical evolution during the formation of molecular clouds in the compression layer behind interstellar shock waves, most C and O appeared to be locked up in CO and solid $\rm H_2O$, respectively.

We have run a model with modified initial abundances for the core stage, where all free carbon is added to CO, borrowing also an appropriate amount of free oxygen, and all remaining free oxygen is added to $\rm gH_2O$ (borrowing an appropriate hydrogen atoms amount from the most abundant $\rm H_2$ species). Compared to our default model, the resulting $\rm H_2CO$ abundance is practically the same in the central area of the core and only a factor of 3 lower at the core edge. Up to $\approx 3 \cdot 10^{4}$~years, $\rm H_2CO$ appears in the gas phase mostly due to the chemical desorption in the reaction $\rm gH + gHCO \rightarrow gH_2CO$, and its abundance in gas grows high, reaching $10^{-8}$ at $10^{4}$~years. After that, $\rm H_2CO$ abundance starts to decrease; however, at $3 \cdot 10^{4}$~years the dominant path to gaseous $\rm H_2CO$ switches to the reaction $\rm CH_3 + O \rightarrow H_2CO + H$, which increases formaldehyde abundance again to $10^{-8}$. The major source of atomic oxygen is the reaction $\rm He^+ + CO \rightarrow C^+ + O + He$, and by $3 \cdot 10^{4}$~years its abundance becomes equal to $3.3 \cdot 10^{-7}$ at the central area of the core. This is a sufficient amount for effective $\rm H_2CO$ production via the reaction with $\rm CH_3$.

\subsection{Reactions that possibly affect \ce{H2CO} chemistry}

The problem of the enhanced $\rm H_2CO$ in the model results could be solved with its effective destruction. The most effective destruction reaction is \ce{H2CO}~+~O $\rightarrow$ CO~+~HCO or \ce{H2CO}~+~O $\rightarrow$ CO~+~OH~+~H, when no activation barrier is present like in \citet{Belloche2014}. In this case, it leads to a decrease in $\rm H_2CO$ abundance for a factor of $\approx 3.0$ and does not change $\rm CH_3OH$ abundance. The lowest \ce{H2CO}~:~\ce{CH3OH} abundances ratio equals $\approx$~2.6 at the $R$ of 8000~au from the core centre, which is still higher than unity. However, other works provide a high barrier of 11--16~kJ~mole$^{-1}$ for this reaction \citep{Herron1988}, with which it cannot proceed at low temperatures characteristic for molecular clouds.

For the reaction $\rm H_2CO + OH \rightarrow HCO + H_2O$, we have a rate of $10^{-11}$~cm$^3$s$^{-1}$ proposed in the OSU database. \cite{Machado2020} suggest an Arrhenius expression for the rate coefficient in the range of $20 < T < 550$~K (Equation~7 in their paper). When extrapolating this to the temperature of 10~K, we obtain $1.26 \times 10^{-11}$~cm$^3$s$^{-1}$, which is similar to our value. The Arrhenius expression by \cite{Ocana2017} for $22 < T < 300$~K (Equation~11 in their paper) implies a rate of $2.57 \times 10^{-10}$~cm$^3$s$^{-1}$ at 10~K, which is an order of magnitude higher than the above-mentioned ones. However, with any of these rate coefficients, the reaction does not affect $\rm H_2CO$ abundance.

\cite{Ruaud2015} suggest Arrhenius constants $\alpha = 10^{-11}$~cm$^3$s$^{-1}$, $\beta = -0.4,\ \gamma = 0$ for the reaction $\rm H_2CO + CN \rightarrow HCO + HCN$. Including this reaction into our network does not influence the $\rm H_2CO$ abundance. The reaction has also been studied at high temperatures \citep{Yu1993}, however, an extrapolation of these results to the cold conditions of prestellar cores is questionable.

Another way to decrease formaldehyde production rate is to spend efficiently its precursor $\rm CH_3$ in other reactions except the formation of $\rm H_2CO$. We tested the reactions $\rm CH_3 + N \rightarrow CH_2N + H$ \citep{Cimas2006, Loison2014}, $\rm CH_3 + HCO \rightarrow CH_4 + CO$ \citep{Hebrard2009} and $\rm CH_3 + OH \rightarrow CH_2 + H_2O$ \citep{Jasper2007}, however, they do not affect $\rm H_2CO$ abundance.

\subsection{Varying the rate of the $\bf CH_3 + O$ reaction}

To obtain the rate constants of the $\rm CH_3 + O$ reaction for a wide temperature range of 50--2000~K, \cite{Yagi2004} performed multistate quantum reactive scattering calculations, as well as classical capture calculations (see their Fig.~7), resulting in a positive temperature dependence. The lowest suggested rate is $0.6 \times 10^{-10}$~cm$^3$s$^{-1}$ at the lowest temperature of 50~K. Implementing this value in our default model leads to decrease of $\rm H_2CO$ abundance by a factor of 1.1 at the core centre.

\subsection{Varying the initial temperature and density}

We varied the initial physical conditions of the diffusive/translucent cloud:
\begin{itemize}
\item initial temperature equals 20~K and is decreasing to 10~K, gas density equals $10^3$~cm$^{-3}$;
\item initial temperature equals 20~K and is decreasing to 10~K, gas density equals $10^2$~cm$^{-3}$;
\item temperature equals 20~K and is not decreasing, gas density equals $10^3$~cm$^{-3}$;
\item temperature equals 20~K and is not decreasing, gas density equals $10^2$~cm$^{-3}$.
\end{itemize}

In all these cases, the abundance of $\rm H_2CO$ either remains the same as in our default model or even increases a little.


\bsp	
\label{lastpage}
\end{document}